\documentclass[sigconf, nonacm]{acmart}

\usepackage{algorithm}          % Algorithms
\usepackage[noEnd=true, indLines=true]{algpseudocodex} % Algorithms
\usepackage{setspace}

\usepackage{mathtools}
\DeclarePairedDelimiter{\ceil}{\lceil}{\rceil}

\usepackage{graphicx}
\usepackage{balance}
\usepackage{amsthm}
\usepackage{bbm}
\usepackage{booktabs, makecell}
\usepackage{multirow}
\usepackage{url}
\usepackage{subcaption}
\usepackage{array}
\usepackage{bbding}
\usepackage{cancel}

\usepackage{cleveref}
\theoremstyle{definition}
\newtheorem{definition}{Def.}[]

\newcommand\vldbdoi{XX.XX/XXX.XX}

\newcommand\vldbavailabilityurl{https://github.com/squash-authors/SQUASH}
\begin{document}
\title{SQUASH: Serverless and Distributed Quantization-based Attributed Vector Similarity Search}

%%
%% The "author" command and its associated commands are used to define the authors and their affiliations.
\author{Joe Oakley}
\affiliation{%
  \institution{University of Warwick}
  \city{Coventry}
  \country{UK}}
\email{J.Oakley@warwick.ac.uk}
\orcid{0000-0003-1926-0665}

\author{Hakan Ferhatosmanoglu}
\authornote{Hakan Ferhatosmanoglu is also with Amazon Web Services. This publication describes work done at the University of Warwick and is not associated with Amazon.}
\affiliation{%
  \institution{University of Warwick}
  \city{Coventry}
  \country{UK}}
\email{Hakan.F@warwick.ac.uk}
\orcid{0000-0002-5181-4712}

\begin{abstract}
Vector similarity search presents significant challenges in terms of scalability for large and high-dimensional datasets, as well as in providing native support for hybrid queries. 
Serverless computing and cloud functions offer attractive benefits such as elasticity and cost-effectiveness, but are difficult to apply to data-intensive workloads. 
Jointly addressing these two main challenges, we present SQUASH, the first fully serverless vector search solution with rich support for hybrid queries.
It features OSQ, an optimized and highly parallelizable quantization-based approach for vectors and attributes.
Its segment-based storage mechanism enables significant compression in resource-constrained settings and offers efficient dimensional extraction operations.
SQUASH performs a single distributed pass to guarantee the return of sufficiently many vectors satisfying the filter predicate, achieving high accuracy and avoiding redundant computation for vectors which fail the predicate. 
A multi-level search workflow is introduced to prune most vectors early to minimize the load on Function-as-a-Service (FaaS) instances. 
SQUASH is designed to identify and utilize retention of relevant data in re-used runtime containers, which eliminates redundant I/O and reduces costs. 
Finally, we demonstrate a new tree-based method for rapid FaaS invocation, enabling the bi-directional flow of data via request/response payloads.
Experiments comparing SQUASH with state-of-the-art serverless vector search solutions and server-based baselines on vector search benchmarks confirm significant performance improvements at a lower cost.
\end{abstract}

\maketitle

% \pagestyle{\vldbpagestyle}
% \begingroup\small\noindent\raggedright\textbf{PVLDB Reference Format:}\\
% \vldbauthors. \vldbtitle. PVLDB, \vldbvolume(\vldbissue): \vldbpages, \vldbyear.\\
% \href{https://doi.org/\vldbdoi}{doi:\vldbdoi}
% \endgroup
% \begingroup
% \renewcommand\thefootnote{}\footnote{\noindent
% This work is licensed under the Creative Commons BY-NC-ND 4.0 International License. Visit \url{https://creativecommons.org/licenses/by-nc-nd/4.0/} to view a copy of this license. For any use beyond those covered by this license, obtain permission by emailing \href{mailto:info@vldb.org}{info@vldb.org}. Copyright is held by the owner/author(s). Publication rights licensed to the VLDB Endowment. \\
% \raggedright Proceedings of the VLDB Endowment, Vol. \vldbvolume, No. \vldbissue\ %
% ISSN 2150-8097. \\
% \href{https://doi.org/\vldbdoi}{doi:\vldbdoi} \\
% }\addtocounter{footnote}{-1}\endgroup
% %%% VLDB block end %%%

\ifdefempty{\vldbavailabilityurl}{}{
\vspace{.3cm}
\begingroup\small\noindent\raggedright\textbf{Artifact Availability:}\\
The source code, data, and/or other artifacts have been made available at \url{\vldbavailabilityurl}.
\endgroup
}

% \input{1-introduction}
% \input{2-quantization-based-indexing}
% \input{3-distributed-filtered-vector-search}
% \input{4-squash-design}
% \input{5-cost-model}
% \input{6-related-work}
% \input{7-experiments}
% \input{8-conclusion}

\section{Introduction}
\label{s:1-intro}

% Paragraph 1 - Nearest neighbor search intro
Nearest neighbor search (\textbf{NNS}) is a fundamental task in a wide range of domains including database systems, information retrieval and recommendation systems.
Recent advances in large language models (LLMs) have introduced new use cases such as retrieval-augmented generation (RAG) which have attracted significant attention. 
NNS involves searching a database of $d$-dimensional vectors to find the nearest vectors to a query, with proximity measured by distance functions (e.g., Euclidean, cosine). 
This becomes computationally challenging as dimensionality and dataset size increase. 
High-dimensional similarity search has been studied using various methods, with quantization and proximity graphs being the dominant approaches in practice. 
While early methods focused on exact search \cite{Weber1998VA, Ferhatosmanoglu2000VAPlus}, 
%often relying on scan-based techniques which evaluate each vector as a candidate. However, 
recent methods prioritize approximate nearest neighbor search (\textbf{ANNS}) \cite{Niu2023ResidualVectorProductQuantization, Gao2024RabitQ, Malkov2020HNSW, Jaiswal2022OODDiskANN, Xu2020, Zhao2022, Gao2023DCOs}, %accepting a trade-off between accuracy and runtime,
trading accuracy for performance to handle large high-dimensional vector embeddings efficiently.

Recent work extends ANNS to support attribute filtering (AF), where the vector similarity is constrained with filters on additional attributes (e.g., price, reviews) \cite{Gupta2023CAPS, Patel2024ACORN}. 
While hybrid ANNS is gaining attention, scalable solutions which efficiently handle complex filters are still lacking.  
In this paper, we introduce \textbf{SQUASH}, a serverless and distributed hybrid vector ANNS system designed as an elastic Function-as-a-Service (FaaS) architecture. SQUASH supports advanced filtering on both continuous and categorical attributes of any cardinality, including complex logical conditions. 

Serverless has been identified as offering a low-cost and performant hosting option for several data-intensive workloads \cite{Muller2020Lambada, Wang2024Starling, Oakley2024FSDInference, Su2024Vexless, Jarachanthan2021AMPS, Oakley2024ForesightPlus, Gillis2021}.
While serverless settings promise cost-effectiveness and elasticity without the overhead of provisioning servers, designing a scalable solution for serverless hybrid vector search involves overcoming significant obstacles. 
These include algorithmic challenges in developing resource-efficient distributed vector indexing for filtered ANNS, as well as systems issues such as memory constraints, cold starts and the runtime limits of FaaS instances. 
Addressing these challenges in a distributed setting is particularly difficult, as we discuss in the paper.

SQUASH is the first serverless and distributed vector search system with native support for hybrid ANNS. 
As part of the indexing approach, we introduce Optimized Scalar Quantization (OSQ), a lightweight and non-uniform quantization method which minimizes FaaS instance load, maximizes parallelism and achieves high compression on vectors, without relying on auxiliary search structures like proximity graphs. 
Our distributed ANNS algorithm retrieves enough vectors to satisfy query predicates in a single parallel pass with minimal communication overhead.
The proposed design enhances both parallelization and compression, which are critical for FaaS environments where high memory usage and limited parallelism hinder scalability.  
In contrast, a graph-based approach would consume substantial memory as well as requiring the partitioning and management of graph structures across instances, increasing overhead and complicating parallelization.  

SQUASH is designed for both vector and attribute data, utilizing dimension-by-dimension quantization to enable effective filtering and parallelism, along with shared segment-based packing of multiple dimensions for improved compression. 
Most existing SQ mechanisms, such as those applied in FAISS \cite{Johnson2021GPUSearch, douze2024faisslibraryivfsq8} and Milvus \cite{Wang2021Milvus}, treat SQ solely as a basic data compressor for individual vector dimensions (e.g., converting 4-byte floats to 1-byte integers). 
In contrast, the proposed non-uniform OSQ approach integrates hybrid search with filtered partition selection, using bitwise operations and accelerated SIMD lookups to prune large portions of vectors early, reducing I/O and avoiding costly distance calculations. 
Unlike prior filtered ANNS approaches, which support basic hybrid search functionality (e.g., single attributes, restricted operators, or low-cardinality tags), we cater for rich filtering for both real-valued and categorical attributes, handling both equality and range queries across multiple attributes with varying selectivity. 

SQUASH achieves fully serverless hybrid search by introducing a number of systems optimizations.
To scale rapidly to thousands of concurrent FaaS instances, a tree-based method is introduced for large-scale invocation with efficient bidirectional data flow via request/response payloads. 
SQUASH identifies retention of relevant data in re-used runtime containers which avoids redundant I/O. Finally, we present a cost model for serverless distributed vector search to inform storage/communication designs and memory/parallelism levels for scalable serverless data-intensive solutions. 
Experiments comparing SQUASH with state-of-the-art serverless vector search solutions and server-based baselines confirm significant performance improvements at a lower cost.
The following summarizes our main technical contributions.
\begin{itemize}
    \item We design OSQ, a distributed quantization-based indexing method for both vector and attribute data. OSQ combines multiple dimensions into shared segments for optimal compression, and enables native filtering support. % and bitwise processing.
    % which enables native filtering and bitwise processing and storage.
    \item A low-bit variant of OSQ is developed for early candidate pruning via fast bitwise comparisons. We observe strong correlations between low-bit, higher-bit and full-precision distance calculations, enabling high recall with low re-ranking requirements.
    % \item The filtered partition selection algorithm guarantees accurate filtered results within a single parallel pass, addressing unique challenges in the distributed setting. 
    \item The multi-stage search pipeline reduces the load on lightweight FaaS instances, significantly pruning via attribute filtering, partition selection and low-bit OSQ. 
    The partition selection algorithm guarantees accurate filtered results within a single parallel pass. 
    In-memory distance look-ups, based on OSQ boundaries, minimize the number of distance calculations. % required by exploiting code repetition in the quantized index.
    \item A tree-based method for synchronous FaaS invocation is designed to quickly scale to thousands of concurrent instances. The multi-level approach and ID selection scheme enables the efficient coordination of parallel workers. A data retention mechanism is also introduced to enable re-use of relevant vector indexes in runtime containers.
    % \item A data retention mechanism is introduced to enable re-use of relevant vector indices in runtime containers.
    \item A cost model and experiments provide insights into the cost/performance characteristics of serverless and distributed vector search.
\end{itemize}

The rest of this paper is structured as follows. Section \ref{s:2-osq} presents OSQ, our quantization-based indexing method. Next, Section \ref{s:3-serverless} describes the fully serverless SQUASH system. Section \ref{s:4-squash-related-work} presents related work. Section \ref{s:5-squash-expts} describes the experimental analysis, before Section \ref{s:6-squash-conclusion} concludes.

\section{Optimizing Scalar Quantization for Serverless Hybrid Vector Search}
\label{s:2-osq}

\begin{figure*}[]
    \centering
    % \begin{subfigure}[t]{0.4\textwidth}
    \begin{subfigure}[t]{0.45\textwidth}
        \centering
        \includegraphics[width=\textwidth]{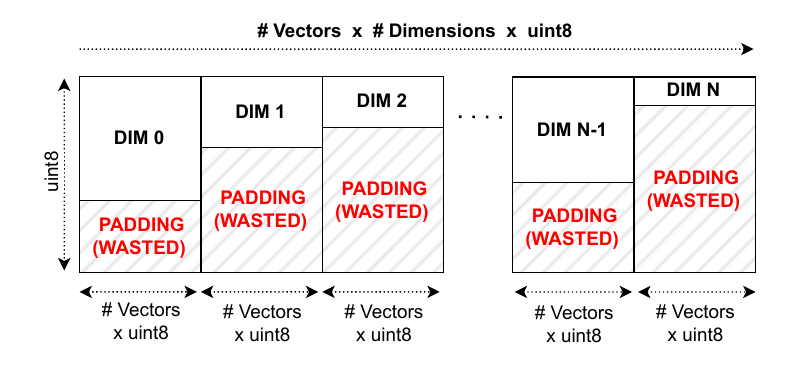}
        \caption{SQ: Individual variables per dimension}
        \label{fig:subfig-osq-a}
    \end{subfigure}%
    % \begin{subfigure}[t]{0.4\textwidth}
    \begin{subfigure}[t]{0.45\textwidth}
        \centering
        \includegraphics[width=\textwidth]{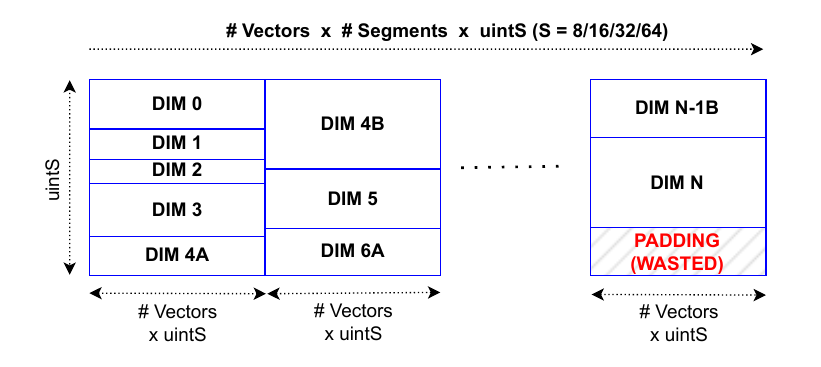}
        \caption{OSQ: Segments shared by multiple dimensions}
        \label{fig:subfig-osq-b}
    \end{subfigure}
    \caption{Comparison of SQ and OSQ storage schemes}
    \label{fig:osq-main}
\end{figure*}

We aim to develop a distributed vector index and architecture design utilizing \textbf{FaaS platforms} (e.g., AWS Lambda \cite{AWSLambdaGeneral}, Azure Functions \cite{AzureFunctionsGeneral}) to deliver large-scale and
low-cost \textbf{hybrid similarity queries} with support for complex predicates across multiple attributes. We seek to provide the ability to scale to zero and support sporadic utilization, avoiding deploying and managing pre-provisioned servers. Our work demonstrates that an optimized \textbf{serverless-centric} index combined with the elasticity of a FaaS-based design provides an efficient and practical solution. 
% --------------------------------------------------------
\subsection{Serverless Hybrid Search Index Design}
\label{ss:2.1-index-design}

%The FaaS setting itself presents additional design challenges.
Our indexing method must maximize accuracy/recall while minimizing latency and cost, particularly for high-recall domains (e.g., 95+\%) and with rich attribute filtering, while satisfying the following requirements: \textbf{1) Parallelizable over FaaS Instances}: It must support significant parallelism in a distributed setting within the limited compute capabilities of individual FaaS instances.
\textbf{2) Low Memory Footprint}: Compact indexes are essential due to constrained memory resources and the lack of local persistent storage across invocations.
\textbf{3) High Recall with Minimal Re-Ranking}: Current ANNS methods either require large in-memory storage of full-precision vectors which is infeasible for FaaS, or incur slow, costly re-ranking due to excessive disk I/O. A scalable solution is needed which does not require substantial memory or suffer from re-ranking delays.
\textbf{4) Suitability for Rich Attribute Filtering}: Current hybrid ANNS methods offer only limited filtering capabilities, with selection based on low-cardinality tags and/or single-attribute, equality-only predicates. In contrast, we aim to handle any number of attributes (of varying data types/cardinalities), as well as complex predicates over any combination of attributes simultaneously. Further, our solution needs to be scalable and handle varying selectivity levels and avoid multiple search passes. %, often caused by insufficiently many vectors satisfying the filter being found on the first pass.

There are four popular categories of ANNS indexing methods---Trees, Hashing, Quantization, and Proximity Graphs---with the latter two dominating the state-of-the-art. Quantization methods, in particular \textbf{Product Quantization (PQ)} and its variants \cite{TuncelFerhatosmanoglu2002VQIndex, Jegou2011PQ, OptimizedProductQuantization, VarianceAwareQuantization, NEURIPS2023-Wang-NHQ} are widely used, both standalone and as part of composite index structures (e.g., with proximity graphs). 
% However, PQ-based solutions are not suitable in our FaaS-based setting, for one key reason; as a standalone index, PQ struggles to provide high recall \cite{douze2024faisslibraryivfsq8}. 
However, despite achieving significant compression, PQ-based solutions struggle to achieve high recall when only the in-memory quantized vectors are considered \cite{Niu2023ResidualVectorProductQuantization, FaissMissingManual, RabitQ2024, Noh_2021_ICCV_PQ_Recall}.
PQ-based methods therefore typically rely on \textit{large amounts of re-ranking on full precision vectors} \cite{RabitQ2024}; this is not viable in the FaaS setting as full-precision vectors are too large to hold in memory and too slow to fetch from disk \textit{in the quantities required}. In contrast, scalar quantization (SQ) can maintain higher accuracy while still achieving considerable compression, which alleviates the need for such extensive re-ranking.

\textbf{Proximity graph}-based ANNS solutions, such as HNSW, offer high recall at low latency. However, whilst various works have studied \textit{attribute filtering} within PG methods \cite{Gollapudi2023FilteredDiskANN, Patel2024ACORN, wang2022navigableproximitygraphdrivennative-NHQ-2, NEURIPS2023-Wang-NHQ, zhao2022constrainedapproximatesimilaritysearchAIRSHIP}, they struggle to offer rich filtering support and assume a level of correlation between vector similarity and attribute similarity, which may not always hold. For example, approaches such as ACORN \cite{Patel2024ACORN} expand the HNSW neighborhoods and/or search scope, based on attribute distributions and assumptions around query predicate selectivity. Not only does this require a priori knowledge of query workloads, it is also focused on single-attribute support; it is challenging to adapt such approaches to cater for multiple attributes simultaneously. SeRF \cite{SeRF2024}, another PG-based approach for range queries, also fails to provide filtering support across multiple attributes simultaneously. 
Incorporating filters can disrupt graph traversal, preventing greedy search from efficiently navigating to relevant nodes.
Further, the ‘decomposition-assembly model’ \cite{NEURIPS2023-Wang-NHQ} whereby hybrid queries are split into two problems, addressed by different indexing solutions, is difficult to apply to PGs. 
In contrast, quantization-based methods are amenable to this approach, making it simpler to adapt them for hybrid search.
Importantly, PG-based approaches also impose \textit{high memory requirements}. For standard HNSW, the full-precision vectors form the nodes of the in-memory graph. 
There are many HNSW variants, aimed at reducing search complexity or memory requirements while maintaining accuracy/recall levels. 
% These include traversing IVF centroids using HNSW as a cluster selection method before brute-force search over full-precision (in-memory) vectors, or using PQ codes as graph nodes with uncompressed vectors also in memory, amongst others. 
For example, PQ codes can be used as graph nodes, with uncompressed vectors also held in memory.
Such solutions still have high memory footprints, or require significant disk-based re-ranking \cite{FaissMissingManual}; neither is desirable in the FaaS setting. 
Parallelization of PG methods is also challenging; partitioning a global graph severs long-range connections, which are essential for traversal efficiency, while pre-partitioning before graph construction isolates boundary nodes by disrupting their true neighborhoods and distorts local neighborhood structures.
We require a highly parallelizable solution which can achieve significant compression without compromising accuracy.
% Parallelizing PG methods is challenging; partitioning a global graph severs long-range connections, disrupting traversal, while pre-partitioning isolates boundary nodes, hindering cross-partition search. We need a highly parallelizable solution that achieves strong compression without sacrificing accuracy.

\textbf{Scalar Quantization (SQ)} is a type of vector compression with several appealing properties that can be enhanced and adapted for serverless hybrid ANNS. First, its scan-based search mechanism on compressed vectors is well-suited for distribution and parallelization, unlike PG-based methods. Due to its ability to achieve higher accuracy with a modest compression ratio (e.g., compared to PQ), it can achieve high recall with low re-ranking requirements, leveraging its contiguous data organization to optimize memory utilization. Since we can design a standalone SQ-based solution that does not require full-precision vectors or auxiliary graph structures to be stored in memory, it imposes only a modest memory footprint. Finally, since the core SQ index structure does not require complex modifications in the filtered case (unlike PGs), it is amenable to be adapted for hybrid search.

However, in recent years SQ has most commonly been considered in its basic form and used as a simple uniform compression technique (e.g., IVF\_SQ8 in Milvus \cite{Wang2021Milvus}/FAISS \cite{douze2024faisslibraryivfsq8}). While classical SQ approaches, such as the VA$^+$-file \cite{Ferhatosmanoglu2000VAPlus}, demonstrate strong performance for large, high-dimensional datasets in centralized computing environments \cite{Echihabi2018HydraSurvey}, they are far from optimal and require modernization to serve as the basis for a distributed, parallel quantization-based indexing scheme with rich attribute filtering, and meeting the serverless FaaS-centric requirements outlined above. Hence, in this section, we introduce our \textbf{Optimized Scalar Quantization (OSQ)} and describe its different components and adaptation at each level of the SQUASH indexing system. An illustration of OSQ's effectiveness for serverless hybrid ANNS over alternatives is summarized in Table \ref{tab:index-features}.

% \newcolumntype{P}[1]{>{\centering\arraybackslash}p{#1}}
% \newcolumntype{M}[1]{>{\centering\arraybackslash}m{#1}}
% \newcommand{\cmark}{\ding{51}}%
% \newcommand{\xmark}{\ding{55}}%

% % \begin{table*}[h]
% \begin{table}[]
% \centering
% \caption{Suitability of several index categories for SQUASH}
% \begin{tabular}{|M{0.32\linewidth}|c|c|M{0.12\linewidth}|M{0.17\linewidth}|}
%     \hline
%      Design Objective & \textbf{OSQ} & PQ & PG (HNSW) & PG (HNSW + SQ/PQ) \\ \hline
%      Parallelizable over FaaS Instances & $\checkmark$ & $\checkmark$ \Checkmark & $\checkmark$ & $\checkmark$ \\ \hline
%      Low Memory Footprint & $\checkmark$ & & & \\ \hline
%      High Recall with Minimal Re-Ranking & $\checkmark$ & & $\checkmark$ & \\ \hline
%      Suitable for Rich Attribute Filtering & $\checkmark$ & $\checkmark$ & & \\ \hline
% \end{tabular}
% \label{tab:index-features}
% % \end{table*}
% \end{table}

\newcolumntype{P}[1]{>{\centering\arraybackslash}p{#1}}
\newcolumntype{M}[1]{>{\centering\arraybackslash}m{#1}}

% \begin{table*}[h]
\begin{table}[]
\centering
\caption{Suitability of several index categories for SQUASH. Brackets indicate partial support.}
% \caption{Suitability of several index categories for SQUASH.}
\begin{tabular}{|M{0.32\linewidth}|c|c|M{0.12\linewidth}|M{0.17\linewidth}|}
    \hline
     Design Objective & \textbf{OSQ} & PQ & PG (HNSW) & PG (HNSW + SQ/PQ) \\ \hline
     % Parallelizable over FaaS Instances & \Checkmark & \Checkmark  & \mbox{}\bcancel{\Checkmark} & \mbox{}\bcancel{\Checkmark} \\ \hline
     Parallelizable over FaaS Instances & \Checkmark & \Checkmark  & (\Checkmark) & (\Checkmark) \\ \hline
     Low Memory Footprint & \Checkmark & & & \\ \hline
     High Recall with Minimal Re-Ranking & \Checkmark & & \Checkmark & \\ \hline
     Suitable for Rich Attribute Filtering & \Checkmark & \Checkmark & & \\ \hline
\end{tabular}
\label{tab:index-features}
% \end{table*}
\end{table}

%----------------------------------------------------------------------
%----------------------------------------------------------------------
\subsection{Optimized Scalar Quantization (OSQ)}
\label{ss:2.2-osq}

%Optimized Scalar Quantization (OSQ) is a serverless-centric indexing method for hybrid vector similarity search.
%While SQ approaches are able to achieve high accuracy, relatively large memory requirements can be incurred if careful optimizations are not made.
Given the inherent resource constraints of the FaaS environment, it is important for the index to achieve strong compression. OSQ introduces a segment-based storage scheme which enables the theoretical compression of SQ to be realized in practice. 
By merging variable-length bit approximations for consecutive dimensions, OSQ reduces bit wastage compared to standard SQ, minimizing padding and ensuring more compact storage. 
OSQ also introduces an extraction scheme to efficiently retrieve individual dimensions using lightweight bit-shift operations and extract the same dimension of all candidate vectors simultaneously.
OSQ is readily applied to both vectors and attributes for quantization and pre-computation of distances. Numerical attributes are quantized in a similar fashion to individual vector dimensions. For categorical variables, we maintain an in-memory mapping from quantized cells to unique attribute values.

% \subsubsection{Segment-based storage}
\subsubsection{Shared Segment-based Data Organization}
We aim to encode an appropriate number of bits for each dimension. The number of bits assigned to a dimension determines the precision of the quantized approximations; each additional bit doubles the number of available quantization cells, which significantly improves representation accuracy. Recent solutions utilizing SQ allocate an equal number of bits (e.g., 8-bits) to each dimension \cite{douze2024faisslibraryivfsq8, ZillizSQ}. 
However, this approach disregards the fact that certain dimensions can be significantly more important than others for distinguishing vectors, e.g., due to their higher variance. 
% particularly when transformations are applied during pre-processing.
This is particularly true when energy-compacting transforms are applied during pre-processing.
While non-uniform bit allocation schemes allow more important dimensions to be represented with higher precision \cite{Ferhatosmanoglu2000VAPlus}, existing methods fail to achieve the theoretical compression in practice due to misalignment of the resulting variable-length bit allocations with available data types. 
Since modern CPU architectures and memory systems are optimized for power-of-two data sizes, libraries only offer 8/16/32/64-bit integers. 
As a result, existing solutions store sub-8-bit dimensions in fixed-length 8-bit variables, wasting bits and leading to larger indexes.

In our approach, the bit patterns of variable-length quantization codes for multiple consecutive dimensions are concatenated into shared $S$-bit storage segments, each of which can be e.g., 8/16/32/64 bits.
This scheme is illustrated in Figure \ref{fig:osq-main}\subref{fig:subfig-osq-b}.
Dimensions are able to overlap multiple segments, and segments may contain some or all of one or more dimensions.
This simple solution makes OSQ optimal in terms of memory requirements. For a given total per-vector quantization bit budget $b$, segment size $S$ and dimensionality $d$, the number of segments required (per vector) under OSQ $G_{OSQ} = \ceil{\frac{b}{S}}$, whereas under standard SQ $G_{SQ} = d$. 
(Illustrative example: $d = 128, S = 8, b = 512$. $G_{OSQ} = 64, G_{SQ} = 128$).

%An important consideration is the distribution of `information' across dimensions.

\begin{figure}[h]%[!h]%
\centering  
\includegraphics[scale=0.4]{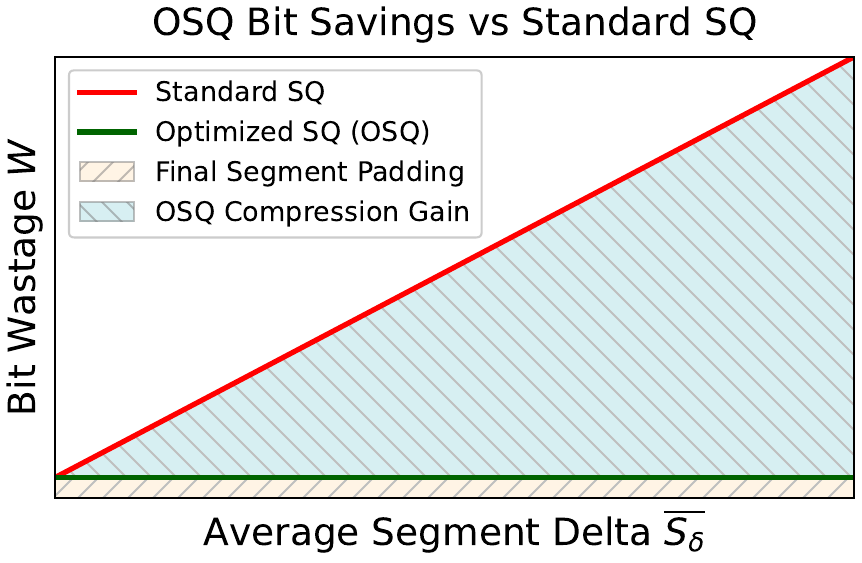}
\Description{Bit Wastage Graph}
\caption{Bit savings under OSQ vs SQ}
\label{fig:bit-wastage-graph}
\end{figure}

Figure \ref{fig:subfig-osq-a} shows the current state-of-the-art SQ data organization under a non-uniform bit allocation of between 0 and 8 bits per dimension.
Since each $B[j]$-length (variable) bit string is stored in an $S$-bit (fixed) representation, SQ suffers from bit wastage in all dimensions where $B[j] \neq S$, as well as the final padding. The degree of wastage $W$ is shown in Figure \ref{fig:bit-wastage-graph} (for a given segment size $S$). We define the segment delta for a given SQ dimension $S_{\delta}(j)$ as the difference between the bits allocated to $j$ and the segment size $S$; $\overline{S_{\delta}}$ is the average for all dimensions. Hence, $W = \sum_{j} S - B[j]$.
In contrast, OSQ incurs the minimal possible bit wastage, i.e., only the padding in the final segment. 
Another benefit of OSQ is the ability to allocate more bits than the segment size (e.g., 9 bits) to a single particularly important dimension. Without OSQ, the size of all segments would need to increase to e.g., 16-bit integer. %, which imposes even more bit wastage.
A combined distance on `fused' dimensions can be computed to further improve the effectiveness of this approach.

To perform the non-uniform bit allocation, bits are iteratively assigned to the dimension with the highest variance; the variance is reduced accordingly after each assignment \cite{gersho2012vector}.
The resulting bit allocations $B$ and quantization cell counts $C$ ($B[j] = k \rightarrow C[j] = 2^k$) are recorded. 
Each dimension $v_{i,j}$ of vector $v_i$ is quantized by identifying its cell within dimension $j$.
Since $j$ contains $C[j] = 2^{B[j]}$ cells, quantized cell positions are encoded in a $B[j]$-length bit pattern. 
The final SQ representation for a single vector concatenates these bit patterns across all dimensions; in OSQ, data is stored in a segment-wise fashion.
A powerful property of the dimension-wise quantization approach is that while the vector space is (non-uniformly) partitioned into $2^b$ (rectangular) cells, individual cells need not be maintained or managed, ensuring the index can remain compact while offering a rich representation.
The dimension-wise approach is also amenable to significant query-time acceleration (e.g., by pre-computing distances from a given query to each quantization cell); we discuss our solution for this in Section \ref{sss:fine-grained-distance-cals}.
%Further work could explore the use of t

% Any more?

\begin{figure*}[]%[!h]%
\centering  
\includegraphics[scale=0.5]{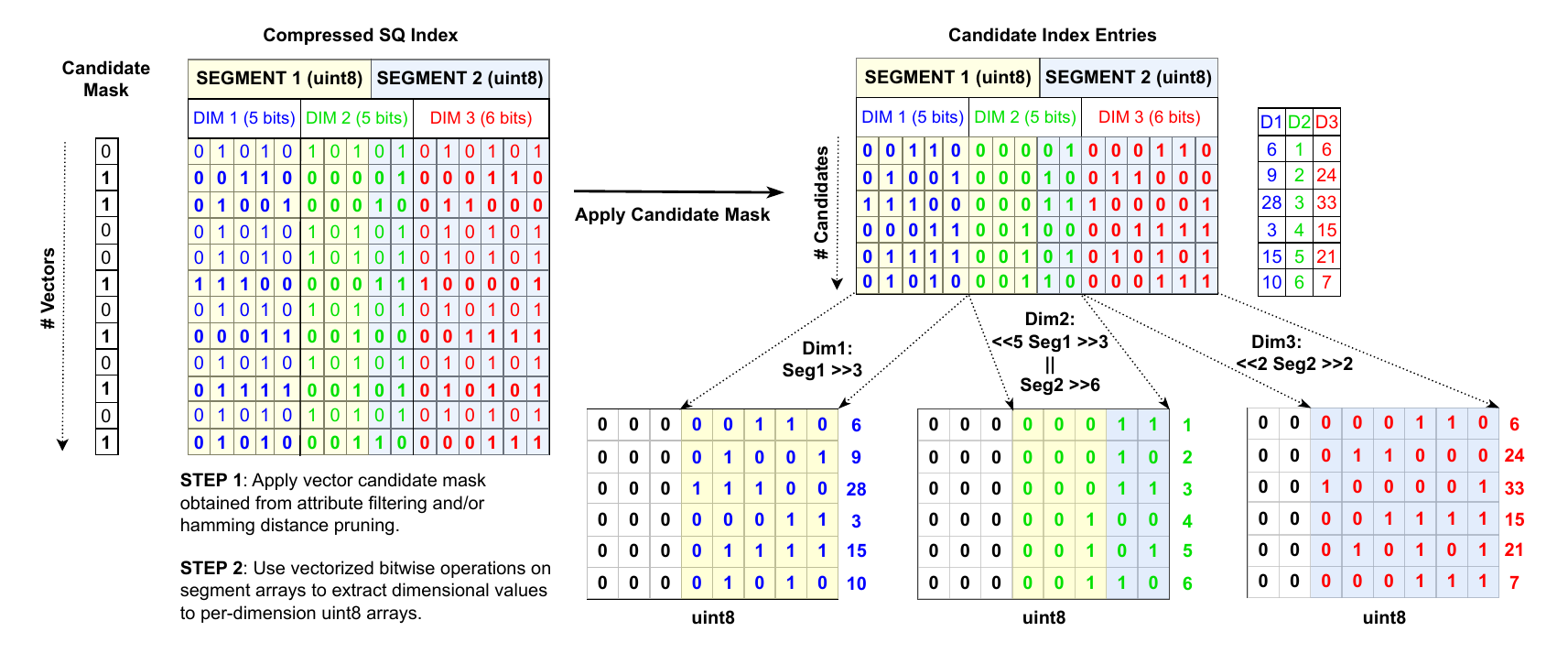}
\Description{Dimensional Extraction Procedure (Bitwise)}
\caption{Illustrative Example of OSQ Dimensional Extraction Procedure, with $S$ = 8}
\label{fig:bitwise-dim-extraction}
\end{figure*}

\subsubsection{OSQ Dimensional Extraction}
In order to access individual dimensions from our OSQ index, we require a mechanism to efficiently extract sub-$S$-bit chunks from an $S$-bit segment. This functionality is required when we seek to compute dimension-wise distances between a query vector and a set of data vectors.
The OSQ extraction scheme utilizes \textit{column-wise} lightweight left/right bit-shift operations for this purpose, as illustrated in Figure \ref{fig:bitwise-dim-extraction}. 
% There are two cases to consider: 1) a dimension is wholly contained in a single segment, 2) a dimension is spread over multiple segments. 
Dimensions may either be wholly contained in a single segment, or spread over multiple segments.
In both cases, we seek to position the relevant bits in the `rightmost' positions, i.e. starting from the LSB.
% , as would be the case for a standard SQ data organization. 
% Could also define an offset, i.e., how many bits away from the left hand start of the segment we are - these aren’t the same value.
In the first case, we perform left-shift operations if the required bits are not on the left side of the segment (e.g., $D_3$ in Figure \ref{fig:bitwise-dim-extraction}), before executing the necessary right-shifts to correctly position the bits. This zeros any earlier bits not relevant to the current dimension.
Extra steps are required to merge the bits from multiple segments (e.g., for $D_2$ in Figure \ref{fig:bitwise-dim-extraction}). The bits from each segment are extracted independently, and positioned in a residue segment $R_i$, with an appropriate offset from the MSB. Let the number of bits of dimension $j$ contained within segment $k$ be denoted as $b_{j,k}$. The 3 bits of $D_2$ in $S_1$ are extracted as in case 1, before being stored in $R_1$ with an offset of ($B[2] - b_{2,1} = 5 - 3 = 2$) bits from the MSB. As the remaining 2 bits of $D_2$ (in $S_2$) are the final bits of $D_2$, they require no offset in $R_2$. A bitwise OR operation is then performed between $R_1$ and $R_2$ in order to produce the final extracted bit representation.
These operations are accelerated with vectorization \cite{NumpyVectorization}, allowing a given dimension to be extracted simultaneously for all vectors. Further, in the hybrid search setting, we typically consider a reduced candidate set after filtering; extraction operations are only performed on rows satisfying the predicate filter (as in Figure \ref{fig:bitwise-dim-extraction}). 
% This further reduces the computational burden of OSQ.

%----------------------------------------------------------------------
%----------------------------------------------------------------------
\subsection{Hybrid Search with Quantized Attributes}
\label{ss:2.3-attribute-filtering}

\begin{definition}[Hybrid Search]
\textit{Given a dataset $D = \{ (v_1, a_1), \dots, (v_N, a_N) \}$ of $N$ vectors $v_i \in$ $\mathbb{R}^d$ with associated attribute data $a_i \in$ $\mathbb{R}^A$, query vector $q = \{v_q , p_q \}$, and limit $k$, find the $k$ vectors in $D$ that are most similar to $q$ which also satisfy predicate $p_q$ (for all attributes). The predicate $p_q$ consists of an operator $m_k$ and one or more operands $n_{k,l}$ for each attribute $k$: $\{(m_0, n_{0,1}, n_{0,2}), \dots, (m_{A-1}, n_{A-1, 1}, n_{A-1, 2}) \}$}.
\label{def:s3-hybrid-search}
\end{definition}

\begin{figure*}[h]%[!h]%
\centering  
\includegraphics[width=\textwidth]{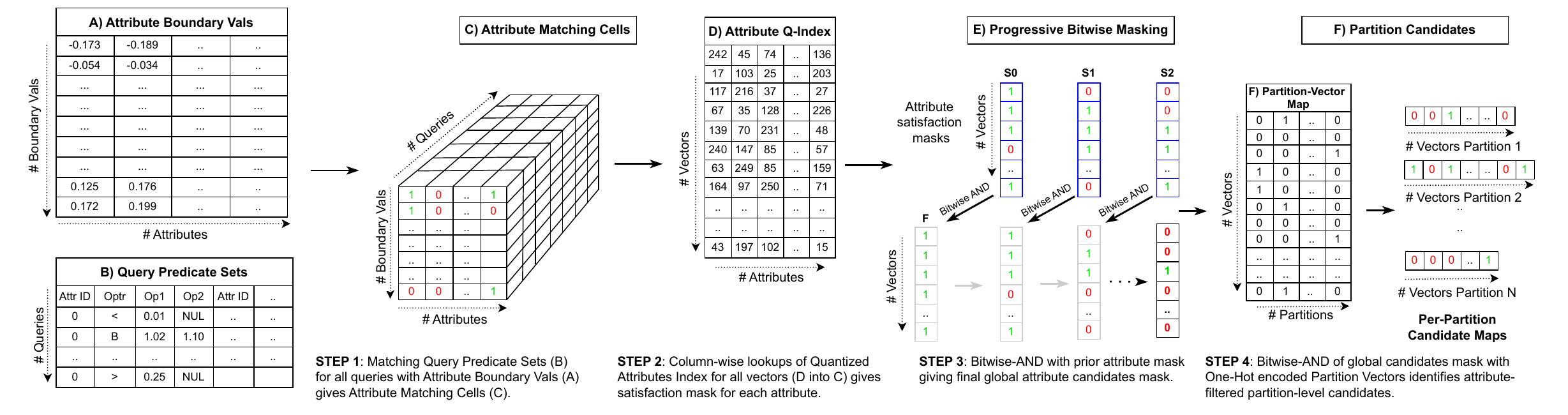}
\Description{SQUASH attribute filtering workflow}
\caption{SQUASH attribute filtering workflow}
\label{fig:squash-attribute-filtering}
\end{figure*}

Our approach for hybrid search, as in Definition \ref{def:s3-hybrid-search}, is illustrated in Figure \ref{fig:squash-attribute-filtering}. We quantize attributes using OSQ in a similar fashion to individual vector dimensions, and employ a cumulative bitwise masking approach to perform filtering.
%advanced indexing \cite{NumpyAdvancedIndexing} as well as 
The final attribute mask is combined with a partition-vector residency map in order to perform distributed filtered search without memory overhead or multiple search passes. The design supports an unlimited number of real-valued or categorical attributes, allowing any combination of equality and range predicates over multiple attributes simultaneously. OSQ minimizes the index size and reduces computational overhead during filtering, making the method highly scalable and suitable for serverless deployments.  This is in contrast to existing filtered ANNS techniques which are limited by one or more constraints: restricting the number and types of attributes, handling only single-attribute queries, or supporting only low-cardinality labels and equality operators. Many also rely on a priori workload knowledge to optimize their filters or indexes. 

\subsubsection{Query Predicate Parsing}
The first step of hybrid query processing is to parse the filter predicate. We present how we support $<, \leq, =, >, \geq$ and `B' (i.e., between, $x \leq y \leq z$) operators. Any combination of these can be provided, and specific attributes can be omitted from the filter. 
% Filter predicates also include 1 or 2 operands (2 are only required for `B').
% For numerical or ordinal categorical attributes, all operators are supported. Only equality operators are supported for nominal categorical attributes.
We maintain an in-memory array of attribute quantization boundary values $V$. 
It has dimensionality ($M + 1, A$), where $M$ is the maximum number of quantization cells in any dimension ($max(C)$) and $A$ is the number of attributes indexed.
At query time, we construct a lookup array $R$, of dimensionality ($M+1, A, |Q|)$, where $|Q|$ is the number of queries to be processed. $R$ contains binary values indicating, for a given query, whether a given quantized cell satisfies the filter for this attribute. % (i.e., that cell `matches' the filter). 
For example, if query $q$ specifies that attribute 0 has value $a_0 < 15$, and $V[:,0] = [0, 5, 10, 15, 20]$, then we set $R[:,0,q] = [1, 1, 1, 0, 0]$. This is illustrated in step 1 of Figure \ref{fig:squash-attribute-filtering}.

\subsubsection{Filter Mask Calculation}
Next, we generate the attribute filter mask, i.e., a binary array indicating the vector IDs that satisfy the filter predicate. Our solution is based on pass/fail bitmaps, and progressively performs efficient bitwise AND operations over all attributes. 
Note that while we present our solution with conjunctive ANDs (i.e., filters for all attributes must hold simultaneously), it is readily extensible to support disjunctive OR predicates.
The attribute filter mask $F$ is a binary array of length $N$, initialized to 1 for all vectors, as none have initially been pruned. We then perform a vectorized lookup into $R[:,0,q]$ (for query $q$) using values from the first column (length $N$) of the Attribute Q-Index, which are held in memory for all vectors. This returns a binary satisfaction array $S_0$, also of length $N$, with 1s in the positions (rows) whose corresponding vectors satisfy the filter predicate for the first attribute. We then perform a bitwise AND between $F$ and $S_0$ to update $F$ (i.e., $F = F \wedge S_0$). Following this step, mask $F$ contains 1s at the positions of vectors which passed the first predicate check, and 0s for vectors which failed. We repeat this process for all attributes, performing successive vectorized lookups/bitwise ANDs to update the mask. Following this, only rows still set to 1 have passed the filter predicates for all attributes, and are carried forward as candidates - the rest are pruned. 
Our approach is also amenable to optimizations based on query workloads. 
For instance, binary filter masks $S_J$ based on popular predicates could be appended to the attribute Q-index and seamlessly swapped in during the progressive bitwise matching phase. 
% This would eliminate the need to identify matching rows for any pre-computed filters.

%----------------------------------------------------------------------
%----------------------------------------------------------------------
\subsection{Multi-Stage Search Routine}
\label{ss:2.4-search-pipeline}

This section describes the multi-stage search routine in SQUASH, in four functional areas, namely coarse partitioner, pre-processing, fine-grained index and post-refinement \cite{douze2024faisslibraryivfsq8}. 

\subsubsection{Coarse Partitioner, Pre-Processing}
To this point, we have presented non-distributed OSQ. We now present its distributed and parallel search adaptation for the serverless FaaS setting.
The coarse partitioning step involves 
constrained clustering to extract balanced partitions for computational load balance in the resource-constrained FaaS environment. 
Alternative balanced partitioning schemes can be utilized, for example fusing vector and multi-attribute similarity information.
Within each partition, we design an OSQ index. We apply an optional unitary, i.e., angle and length-preserving, transformation in order to decorrelate dimensions and 
compact energy which boosts the performance of our non-uniform bit allocation. In our implementation, we used the Karhunen-Loève Transform (KLT) for this purpose.
% This also improves the quality of our non-uniform bit allocation, as its 
We transform each partition independently which is not only more efficient and parallel, but it also captures the local data distribution more accurately; the distance-preserving nature of unitary transforms ensures queries are still processed correctly when results are combined from multiple partitions.
Finally, we perform efficient one-dimensional K-means clustering to design optimal scalar quantizers based on the data distribution \cite{LloydsAlgorithm1982}. 

% Rather than performing a uniform SQ (e.g., equally spaced cells), we execute Lloyd’s algorithm, which designs optimal scalar quantizers for each dimension based on the data distribution. 

\renewcommand{\algorithmicrequire}{\textbf{Input:}}

\begin{algorithm}[h]
\begin{algorithmic}[1]

\setstretch{1.1}
\caption{Filtered Partition Ranking and Selection}
\label{alg:squash-cluster-ranking}

\Require Filter Mask $F$, P-V Map $P_V$, Centroids $C$, Centroid Distance Threshold $T$, Top-K target $k$, Queries $Q$
\State $P_{Q} \gets \{\}$
\ForAll{$q \in Q$}
    \State $Q_{cands}, C_{dists} \gets 0, []$ % Change symbols? 
    \ForAll{$c \in C$} \Comment{\textit{Distances to each partition centroid}}
        \State $C_{dists}[c] \gets \Call{CalcDistance}{q, c}$
    \EndFor
    \ForAll{$p \in $ \Call{ArgSort}{$C_{dists}$}}
        \If{($C_{dists}[p] > T) \wedge (Q_{cands} \geq k$)} %\Comment{\textit{Both termination conditions met}}
            \State ${\bf break}$
        \EndIf 
        \State $p_{cands} \gets $ \Call{FilterPartitionVectors}{$F, P_V, p$} %\Comment{\textit{Vectors in $p$ passing filter $F$}}
        \If{$|p_{cands}| > 0$}
            \State $P_{Q}[p] \gets P_{Q}[p] \cup (q, p_{cands})$ %\Comment{\textit{Add $q$ to query list for partition $p$}}
            \State $Q_{cands} \mathrel{+}= |p_{cands}|$
        \EndIf
    \EndFor
\EndFor\\
\Return $P_{Q}$ \Comment{\textit{Required query visits for each partition}}

\end{algorithmic}
\end{algorithm}

\subsubsection{Filtered Partition Ranking and Selection}
At query time, we first carry out the attribute filtering workflow described in Section \ref{ss:2.3-attribute-filtering}. The attribute satisfaction mask $F$ (output of Figure \ref{fig:squash-attribute-filtering}, Step 3) is calculated globally (i.e., across all partitions).
Next, we determine which partitions to visit to answer a given query. 
This should be achieved in a single pass (i.e., performing per-partition processing only once per query) in order to avoid processor re-invocation.
% minimize I/O, runtime and costs.
\textit{We guarantee that should they exist globally, at least $k$ vectors satisfying the filter predicate will be identified.} For all visited partitions, we consider \textit{all} vectors passing the filter as candidates. %, thus ensuring we find the true nearest neighbors.
% while still guaranteeing the return of sufficiently many vectors which pass the filter predicate. 
% This is challenging in the distributed setting, where a first attempt identifying fewer than $k$ satisfactory vectors requires re-invocation of some or all of the per-partition processors.
\begin{equation}
    T = 1 + \frac{\sigma_{\mu}}{\mu_{\mu}} + \beta \sqrt{d}
    \label{eqn-cdf-calculation}
\end{equation}

Our goal is to visit the minimal set of partitions while satisfying both of the following: 1) at least $k$ vectors passing the filter will be considered, 2) all partitions whose centroids are within a multiplicative distance factor $T$ of the nearest (to the query) are visited (to ensure high recall).
To compute $T$ for a dataset, we first calculate the pairwise distance matrix between all vectors and centroids, before deriving the ratio matrix $R$ by dividing each value by the home centroid distance (home ratios are 1, and others >1). 
The row-wise means ($\mu_R$) and standard deviations ($\sigma_R$) of $R$ are then computed.
In Equation \ref{eqn-cdf-calculation}, $\mu_{\mu} = $ \textsc{mean}$ (\mu_{R}) $ and $\sigma_{\mu} = $ \textsc{mean}$ (\sigma_{R}) $ are the means of the row-wise means/standard deviations of $R$, respectively. $d$ is the dimensionality, while $\beta$ is a small value (e.g., 0.001) which can be tuned based on recall requirements. 

Algorithm \ref{alg:squash-cluster-ranking} shows our partition selection approach. 
We first initialize a dictionary $P_Q$ of queries to be issued to each partition (L1).
% To perform partition ranking for a given query, distances from the query vector to each partition centroid are first calculated (L4-5). 
For each query, distances to each partition centroid are calculated (L4-5). 
Partitions are then considered in ascending distance order (L6). A compact in-memory bitmap $P_V$ of the vectors resident in each partition is maintained.
The \textsc{FilterPartitionVectors} function (L9) uses $P_V$ and the attribute filter mask $F$ to find the vector indices \textit{in a given partition} $p$ satisfying the predicate. 
% If matching vectors are found, we add a bitmap representation to the query metadata for that partition, indicating the \textit{local} indices of the candidate vectors for this query (L11). 
If matching vectors are found, a bitmap representation is added to $P_Q[p]$, indicating \textit{local} candidate indices (L11).
This allows the per-partition processors to prune all non-passing vectors. 
Partitions are continually considered until both 1) $T$ has been reached, and 2) at least $k$ vectors have been identified (L7).
As an optional step in the batch setting, we can issue additional queries to partitions which few queries are currently searching. 
Each such partition is assigned the queries which it was most narrowly previously pruned from (i.e., where it had the first centroid distance above the threshold), until a balanced workload is achieved.
% until a balanced query workload is achieved across all partitions.

% \subsubsection{Pruning via Binary Quantization}
\subsubsection{Fast Pruning via Low-Bit OSQ}
Having reduced the search scope via attribute filtering and partition selection, we wish to quickly prune a significant portion of the remaining local (partition-level) candidates with minimal accuracy loss, allowing full distance calculations to be avoided for as many vectors as possible. This is particularly important when the filter predicate is not highly restrictive, to reduce the load on resource-constrained FaaS instances.
Therefore, in addition to the primary index, we also build a low-bit OSQ index (also in-memory), which assigns a single bit to each dimension and can be leveraged for early candidate pruning via fast bitwise comparisons.
The low-bit OSQ scheme utilizes binary quantization, in conjunction with our segment-based storage scheme, to consolidate the 1-bit representations for $S$ dimensions into each $S$-bit segment.
We first standardize the data, before thresholding vectors around 0 to determine 0/1 mappings. Following this, binary values are stored into segments via OSQ.
% packed into a bitmap, in order to minimize memory and I/O overheads.

We compute (binary quantized) query-to-vector Hamming distances for all local candidates. We observe that Hamming distance using OSQ-based binary vectors
approximately maintains the same relative ordering as lower bound (LB) Euclidean distance for high-dimensional data.
This can be intuitively explained by the shared boundary-based proximity and dimensional independence properties.
Experiments (not shown) indicate strong correlations between the query-to-vector Euclidean distance and the Hamming distances computed based on binary OSQ vectors on a range of datasets. 
The binary OSQ-based distances 
serve as a lightweight bitwise (albeit coarse) approximation for the final results.
The Hamming distance between two binary vectors $x$ and $y$ of equal length $n$ is defined as the number of positions at which corresponding bits differ. 
% If $x$ and $y$ are binary vectors of length $n$, 
The Hamming distance $d_H(x, y)$ is:
\begin{equation}
    d_H(x, y) = \sum_{i=1}^{n} \mathbbm{1}(x_i \neq y_i)
    \label{eq-hamming-dist}
\end{equation}
Where $\mathbbm{1}(x_i \neq y_i)$ is an indicator function that equals 1 if $x_i \neq y_i$, and 0 otherwise.
The percentage cutoff $H_{perc}$ (i.e., the proportion of the best vectors in ascending Hamming distance order to retain) can be tuned according to the approximation tolerance of the user.

\subsubsection{Fine-Grained Distance Calculations}
\label{sss:fine-grained-distance-cals}

Using the primary OSQ index, we approximate the distance from a query $q$ to a given vector by identifying the distance from $q$ to the nearest edge of the high-dimensional cell containing that (quantized) vector; this is a lower bound (LB) distance calculation.
This is calculated as the square root of the sum, over all dimensions, of the squared distances to the nearest boundary values \cite{Weber1998VA}; i.e., the right boundary if $j < cell(q[k])$, the left boundary if $j > cell(q[k])$, and if $j = cell(q[k])$, then the LB distance is 0 for that dimension.
Since the query is un-quantized, this is an asymmetric distance calculation (ADC) \cite{Jegou2011PQ}.
We only calculate LBs for candidates not pruned thus far, and candidates are ranked in ascending LB distance order.

% SQ-based methods calculate LB distances by summing per-dimension query-to-boundary distances for each candidate vector.
Importantly, we note that many vectors are quantized to the same cell in a given dimension.
In existing SQ-based methods, this leads to many redundant computations when processing multiple candidates with shared per-dimension quantized values. 
To address this, we construct an in-memory ADC lookup table $L$ of dimensionality ($M + 1, d$) for each query $q$, where $M$ is the maximum number of quantization cells in any dimension ($max(C)$).
This enables each squared `query dimension' to `dimension boundary value' distance to be calculated only once.
% This enables the distances from the un-quantized query values to the relevant quantization cell boundaries to be calculated only once. 
Building $L$ requires only $(\sum_j C[j]) - 1$ calculations, and is performed extremely efficiently with vectorized operations. 
As described above, $L[j,k]$ contains the distance from $q[k]$ (un-quantized) to the \textit{nearest} quantization boundary value in dimension $j$. 
We then use the lookup to calculate LB distances for all unpruned candidates by performing `advanced indexing' \cite{NumpyAdvancedIndexing} operations; i.e., using the in-memory quantized values (only for remaining local candidates) to index into $L$ for all dimensions, before performing row-wise sums to give final per-vector LB distances.

\subsubsection{Post-Refinement and Result Aggregation}

As an optional step, random reads to disk can be performed to fetch un-quantized vector data, and compute full-precision distance calculations. This would boost recall and improve the ordering of the final result set. When doing so, $R \cdot k$ ($R > 1$) records are retrieved, enabling the discovery of any true nearest neighbors with LB distances just outside the top-$k$. 
Importantly, OSQ enables the use of small $R$ values, typically 2, in contrast to many PQ/PG approaches which require substantial re-ranking, e.g., $R > 100$. This low re-ranking requirement is important in making OSQ suitable for serverless ANNS.
As a final step, we perform an MPI-style reduce operation, where per-partition `local results' are merged at the global level. This entails performing a merge sort of the result sets from each per-partition processor to determine the global top-$k$.
\section{SQUASH Serverless Design}
\label{s:3-serverless}

% \begin{figure*}[]%[!h]%
% \centering  
% % \includegraphics[scale=0.4]{figs/arch6_cropped.pdf}
% % \includegraphics[scale=0.45]{figs/arch6_cropped.pdf}
% \includegraphics[width=\linewidth]{figs/arch6_cropped.pdf}
% \Description{SQUASH high-level architecture}
% \caption{SQUASH high-level architecture}
% \label{fig:squash-overall-architecture}
% \end{figure*}

\begin{figure}[]%[!h]%
\centering  
\includegraphics[scale=0.45]{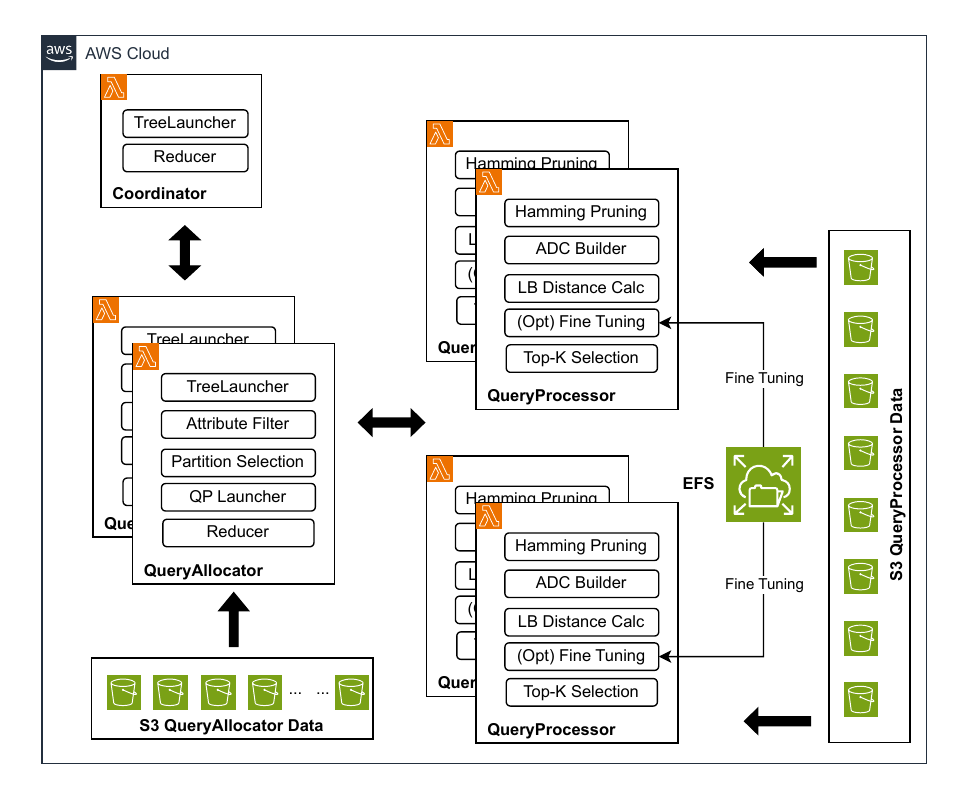}
\Description{SQUASH high-level architecture}
\caption{SQUASH high-level architecture}
\label{fig:squash-overall-architecture}
\end{figure}

\textbf{SQUASH} is deployed on commodity Function-As-A-Service (FaaS) platforms. 
Whilst our solution is cloud provider-agnostic, we present our design using AWS services, in particular AWS Lambda due to its best-in-class performance \cite{Rifai2021, ServerlessSurvey2022}. 
Equivalent solutions can readily be built on other cloud platforms. The high-level SQUASH architecture is shown in Figure \ref{fig:squash-overall-architecture}.

\subsection{SQUASH Run-Time Entities}
The run-time (i.e., query-time) system splits the attributed vector search task over three key components. %: \textit{Coordinator}, \textit{QueryAllocator} and \textit{QueryProcessor}.
% which are responsible for parsing user input, performing attribute filtering/cluster selection, and the partition-level vector search, respectively.
The first is the Coordinator (\textbf{CO}), which is the hub of SQUASH. It parses user input, initiates the FaaS invocation tree, and returns consolidated results. The next component, the QueryAllocator (\textbf{QA}), operates in a query-parallel fashion. Each QA begins by synchronously launching a set of child QA instances.
It determines its own role in the invocation tree (see Section \ref{sss:4-tree-launch}, i.e., branch or leaf), and how many child instances it should launch, based on the size and depth of the tree. It launches each of its child FaaS instances on a separate thread, and then continues with its own processing tasks. 
These consist of the attribute filtering and partition selection procedures described earlier. 
For settings involving multiple queries, it batches together the relevant queries for each partition, builds the required FaaS payloads, and synchronously launches a set of QueryProcessor (\textbf{QP}) instances, one per partition visited, each on a separate thread. 
% Once QP responses are received, the QA merges the results to obtain the global top-$k$ results for its query set, and consolidates its results with those received from its child QA instances, before returning them to its parent QA (or the CO). 
Once QP responses are received, the QA merges the results to obtain the global top-$k$ results for its query set, and returns them to its parent QA (or the CO), together with any results it received from child QAs. 
The QP component performs the per-partition processing. Having received query metadata (possibly for many queries) from its calling QA, it performs the low-bit OSQ pruning and fine-grained distance calculation steps, as well as the optional post-refinement stage. It then packages the final top-$k$ results for its query set and returns them to its parent QA instance.

\subsection{Data Retention Exploitation (DRE)}

\begin{figure}[h]%[!h]%
\centering  
\includegraphics[scale=0.3]{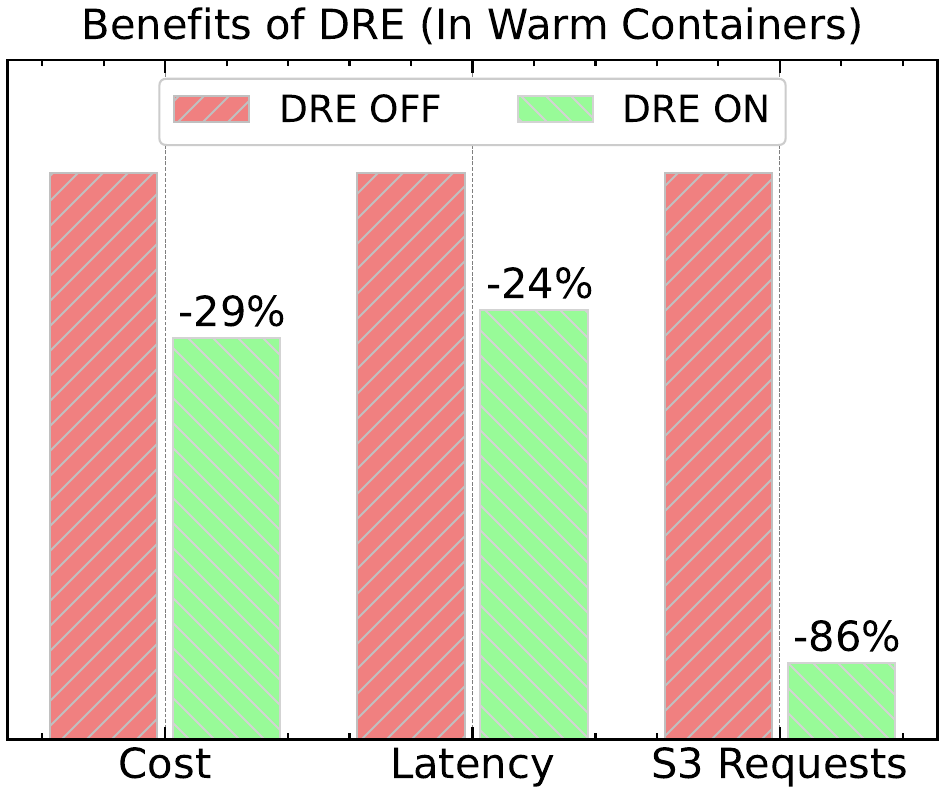}
\Description{Container Re-Use Benefits}
\caption{Cost, latency and S3 request reduction with DRE. Tested with SIFT1M, $N_{QA}=84$}
\label{fig:dre-benefits}
\end{figure}

FaaS-based solutions can benefit greatly from `warm starts' (i.e., where a previously used function container/execution environment can be re-used for a subsequent execution), due to the reduced invocation latency it provides. However, existing data-intensive FaaS systems do not distinguish between warm and cold starts in their processing logic, often leading to significant repetition of I/O operations (e.g., fetching data from external storage). We introduce Data Retention Exploitation (\textbf{DRE}) to enable subsequent invocations to entirely avoid external I/O associated with reading OSQ index files, significantly improving performance and reducing costs as shown in Figure \ref{fig:dre-benefits}.

% We further exploit the use of warm containers by using 
DRE utilizes `singleton classes' to store key data (e.g., vector and attribute indexes) in global areas which are retained across cloud function invocations due to container re-use. AWS Lambda function executions have an initialization (\textsc{Init}) phase, during which any static code (defined outside the `handler' entry point) is invoked.
% Any static code is executed in the \texttt{INIT} phase. 
Each QA/QP instantiates a singleton object (a class which allows only a single instance to exist) as static code. When the QA/QP handler functions are invoked (\textsc{Invoke} phase), before downloading their data from object storage, they first check if the relevant data is already held in the singleton object. If so, and the dataset details match, the fetch request is avoided and the global data is re-used. Otherwise the file is downloaded from storage, and its data items placed in the singleton object, so that further invocations reaching the same runtime container can reuse it. In the case of the QP, the existence of a differently named function for each data partition (\textit{squash-processor-0, squash-processor-1 etc.}) means a QP instance can safely re-use partition-level index data read into memory in a previous invocation, as it is sure to relate to the same partition.

Separately, we also developed an optional lightweight \textit{result} caching solution which saves results from earlier queries and avoids re-processing repeated requests. % requests which have previously been answered. 
This is an additional feature for real-world use cases, where a subset of queries may have recurring patterns. We disable this feature by default and use it only for a specific subset of experiments described in Section \ref{ss:7-caching}.

\subsection{Tree-based FaaS Invocation}

\renewcommand{\algorithmicrequire}{\textbf{Input:}}

\begin{algorithm}[h]
\begin{algorithmic}[1]

\setstretch{1.1}
\caption{Tree-based FaaS Invocation}
\label{alg:squash-lambda-launch}

\Require Branching Factor $F$, Level $l$, Max Level $l_{max}$, $id$
\State $N_{QA}$ $\gets F \frac{1 - F^{l_{max}}}{1 - F}$
\If{$id == -1$} \Comment{Coordinator}
    \State $J_S \gets \ceil*{\frac{N_{QA}}{F}}$ % js or j? Or another letter... skip?
    \For{$i\gets0$ to $F$}
        \State \Call{Invoke}{$id = id + (i \times J_S) + 1$, $l = l + 1$} \Comment{Sync. invoke} %\Comment{Synchronous FaaS invocation}
    \EndFor
\ElsIf{$l_{max} - l \geq 1$} \Comment{Internal QA}
    \State $P_{S} \gets \ceil*{\frac{N_{QA}}{F}}$ \Comment{$P_S$: Parent skip}
    \For{$k\gets0$ to $l$} \Comment{Step down tree}
        \State  $J_S \gets \ceil*{\frac{P_{S} - 1}{F}}$
        \State $P_{S} \gets J_S$
    \EndFor
    \For{$i\gets0$ to $F$}
        \State \Call{Invoke}{$id = id + (i \times J_S) + 1$, $l = l + 1$} \Comment{Sync. invoke} %\Comment{Synchronous FaaS invocation}
    \EndFor
% \EndIf\\
\EndIf
% \Return id

\end{algorithmic}
\end{algorithm}

\begin{figure}[h]%[!h]%
\centering  
\includegraphics[scale=0.5]{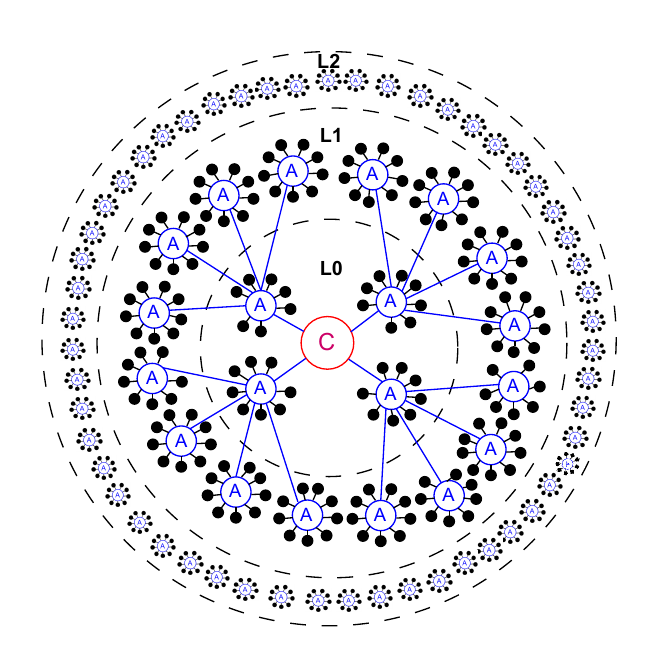}
\Description{SQUASH tree-based Lambda invocation scheme}
\caption{Tree-based FaaS invocation scheme. Blue circles represent QAs, while black circles correspond to QPs.}
\label{fig:squash-lambda-invocation}
\end{figure}

\label{sss:4-tree-launch}
In a naïve parallel FaaS solution, the CO would sequentially invoke all the required QAs. 
However, not only could this require the CO to consecutively perform hundreds of threaded invocations, but it could also lead to a bottleneck at the CO, as all results would need to be returned to this single function instance. 
Instead, we present a \textbf{tree-based multi-level invocation structure}, which offloads most of the invocation/response gathering workload to the allocators themselves. 
We launch multiple levels of QAs, each of which is responsible for invoking and gathering results from all QAs in the sub-tree rooted below itself. 
This is similar to the invocation structure for a fully serverless DNN inference mechanism \cite{Oakley2024FSDInference}, but 1) our scheme is based on synchronous invocation, thus enabling bi-directional data flow via request/response payloads, without an intermediate storage/synchronization layer, and 2) our design targets and is evaluated with significantly higher parallelism levels. 
Our invocation scheme is illustrated in Figure \ref{fig:squash-lambda-invocation} and Algorithm \ref{alg:squash-lambda-launch}. 
We categorize nodes as being one of three types; CO (tree root), \textit{internal} QA (all levels below the root but above the leaves) and \textit{leaf} QA (which invoke no children). 
This approach can flexibly scale to hundreds or thousands of concurrent instances, while requiring only a small number of invocations by each function, reducing the risk of I/O bottlenecks at highly connected source nodes.
We parameterize the solution by a `branching factor' $F$ (how many QAs a node should invoke) and the maximum number of levels $l_{max}$. 
The algorithm begins with the CO ($id = -1$, $l = 0$). It defines a `jump size' $J_S$, indicating the ID gaps it should leave between consecutive children at the next level. 
This maintains the property that for two nodes invoked by the same parent, with IDs $x_i$ and $x_{i+J_S}$, the sub-tree rooted at node $x$ contains all IDs y such that $x_i < y < x_{i+J_S}$. 
% This ensures that each node knows which child IDs it should expect information to be returned from.
This enables each node to know the child IDs to expect information to be returned from.
Upon invocation, each QA runs Algorithm \ref{alg:squash-lambda-launch} to determine which child QA (if any) it needs to invoke, before building the required payloads and performing the FaaS invocation requests. This scheme can result in the rapid launch of hundreds of parallel QAs (and hence thousands of QPs), while avoiding response gathering bottlenecks at single functions. 
% The solution also enables the coordinator to take on the role of an allocator (avoiding additional QA invocation time/costs), if the query volume is low.

\subsection{Task Interleaving for Multi-Query Optimization}
The QA can perform task interleaving to increase utilization while waiting for child QPs to return results. This enables an effective overlap of `communication' (i.e., waiting for child QPs to return results via synchronous response payloads) with computation, in multi-query workloads.
While synchronous invocation would normally be blocking, we instead invoke QAs/QPs via background threads, freeing up the calling FaaS instance for other work. In particular,
% we prepare query metadata (e.g., attribute filtering, partition centroid-query similarity ranking and selection) 
we perform attribute filtering and partition selection 
whilst waiting for QP responses from the previous query/batch. We then capture the per-partition results from QPs, and perform the required merge sort to produce global results. 
% (these steps form the `reduce' phase in Figure \ref{fig:squash-pipeline}). 
This process (\textit{prepare}, then loop: \{\textit{invoke, prepare, reduce}\}) can be repeated for multiple successive queries/batches as required.

\subsection{Cost Model for Serverless Vector Search}
\label{s:5-squash-cost-model}

In this section, we formalize the cost model of SQUASH. 
It should be noted that while we henceforth refer to AWS terminology and pricing models, the design principles are cloud-provider agnostic. %, and similar pricing models apply to other cloud platforms. 
A detailed model is important due to the cost-to-performance tradeoffs of serverless, and to show the viability of SQUASH when compared with alternative serverless vector search solutions.
While previous work~\cite{Muller2020Lambada, Oakley2024FSDInference} has evaluated the costs of serverless object storage-based exchange operators and ML inference, ours is the first to focus on highly parallel vector search.
The cost model is comprised of three main components; AWS Lambda compute costs, S3 (object storage) retrieval costs and EFS (file system) read costs.
\begin{gather}
    C_{Total} = C_{\lambda} + C_{S3} + C_{EFS} \label{eq-cost-total}\\
    C_{\lambda} = C_{Invoc} + C_{Run}\label{eq-cost-lambda}\\
    C_{Invoc} = (N_{QA} + N_{QP} + 1)\cdot C_{\lambda(Inv)}\label{eq-cost-lambda-invoc}\\
    C_{Run} = (M_{QA}\sum_{i=1}^{N_{QA}}T_{A_i} + M_{QP}\sum_{i=1}^{N_{QP}}T_{P_i} + M_{CO}T_{CO}) \cdot C_{\lambda(Run)}\label{eq-cost-lambda-run}
\end{gather}
% Lambda compute costs are made up of two components, namely invocation and runtime. 

In Equations \ref{eq-cost-lambda-invoc} and \ref{eq-cost-lambda-run}, $N_{QA}$ and $N_{QP}$ are the number of QAs and QPs respectively (the CO is the extra 1). 
This is multiplied by $C_{\lambda(Inv)}$, the static cost per Lambda invocation. 
$M_{QA}$, $M_{QP}$ and $M_{CO}$ reflect the memory (in MB) assigned to QAs, QPs and the CO, respectively. 
At present on AWS Lambda, $128 \leq M_X \leq 10240$. 
We sum the per-allocator/processor runtimes $T_{QA_{i}}$ and $T_{QP_{i}}$, as well as the coordinator runtime $T_{CO}$, and multiply by the respective memory allocations to obtain the total number of MB-seconds utilized by SQUASH. This total is multiplied by $C_{\lambda\text(Run)}$, the cost per MB-second of Lambda runtime\footnote{https://aws.amazon.com/lambda/pricing}. %\cite{FN-AWS-Lambda-Pricing}. 
Since AWS Lambda vCPU allocation is proportional to the amount of memory, this introduces an inherent cost-to-performance trade-off \footnote{https://docs.aws.amazon.com/lambda/latest/operatorguide/computing-power.html}. %\cite{FN-AWS-Lambda-Memory-Compute}.
\begin{gather}
    C_{S3} = LC_{S3(Get)}\label{eq-cost-s3}\\
    C_{EFS} = (SR_{Size})C_{EFS(Byte)}\label{eq-cost-efs}
\end{gather}

We utilize two different storage solutions for SQUASH, to maximize performance while reducing costs. 
Specifically, we use object storage (AWS S3) for the OSQ index files for QA/QPs (e.g., quantized vectors/attributes, quantization boundary values, low-bit OSQ index), while we use a cloud-based network file system (AWS EFS) for the full-precision vectors.
We used object storage for the OSQ index files as these reads are (comparatively) large, and S3 does not charge based on the quantity of data transferred to AWS Lambda.
We utilized EFS for the full-precision vectors due to its sub-millisecond random read latencies. While large/frequent reads from EFS can be expensive, we incur low costs due to the low re-ranking requirements of OSQ.
In Equations \ref{eq-cost-s3} and \ref{eq-cost-efs}, $L$ corresponds to the number of GET requests performed, $S$ reflects the number of random reads performed, while $R_{Size}$ indicates the size of a single full-precision vector on disk. $C_{S3(Get)}$ and $C_{EFS(Byte)}$ are the costs per S3 GET request and EFS byte read via Elastic Throughput reads, respectively. Note that for SQUASH and all baselines, we only consider costs incurred by querying, rather than ongoing storage.
\section{Related Work}
\label{s:4-squash-related-work}

Solutions for vector similarity search (nearest-neighbor search (NNS) / approximate nearest neighbor search (ANNS)) fall into categories including hashing \cite{db-lsh-hashing-2, AndoniOptimal-Hashing-3, ANN-Hashing-5, IntelligentProbingLv-Hashing-8, ParkNeighbor-Hashing-9, PM-LSH-Hashing-11}, trees \cite{HouleRankBasedTree, LuVHPHypersphere, Muja2014ScalableNN, SilpaAnan2008OptimizedKDTree}, quantization \cite{ferhatosmanoglu2001approximate, TuncelFerhatosmanoglu2002VQIndex, Ferhatosmanoglu2006VAPlusApprox, Jegou2011PQ, OptimizedProductQuantization, Wang2020DeltaPQ, VarianceAwareQuantization, Aguerrebere2023, aguerrebere2024locallyadaptivequantizationstreamingvector, PQCacheLocality2015, RabitQ2024} and proximity graphs (PG) \cite{FuFastApproximate, Gollapudi2023FilteredDiskANN, NEURIPS2019-DISKANN, Jaiswal2022OODDiskANN, MALKOV201461, Malkov2020HNSW, singh2021freshdiskannfastaccurategraphbased, Zhao2020SONGGPU}, with a multitude of variations/combinations of these themes. 
Scalar quantization methods generate highly compressed and parallelizable representations with lower reproduction errors \cite{douze2024faisslibraryivfsq8}. 
Approaches based on scalar and vector quantization, such as the VA$^+$-file \cite{Ferhatosmanoglu2000VAPlus}, VQ-Index \cite{TuncelFerhatosmanoglu2002VQIndex}, and Product Quantization (PQ) \cite{Jegou2011PQ} are standalone solutions, and are also applied to compress the vectors within coarse index structures such as IVF \cite{TuncelFerhatosmanoglu2002VQIndex, douze2024faisslibraryivfsq8} and proximity graph (PG)-based approaches, such as HNSW \cite{Malkov2020HNSW, FaissMissingManual} and DiskANN \cite{NEURIPS2019-DISKANN}.

\textit{Attributed} vector similarity search based on non-PG solutions has primarily been addressed through two methods: pre-filtering and post-filtering. 
Pre-filtering, used by systems such as AnalyticDB-V \cite{Wei2020AnalyticDBV} and Milvus \cite{Wang2021Milvus}, first searches a separately maintained attribute index using a query predicate supplied with the query feature vector; the filtered candidate list is then used to reduce the scope of the vector similarity search. 
Post-filtering, seen in approaches such as VBASE \cite{Zhang2023VBASE}, first performs the vector similarity search and then prunes the results using the attribute index. 
In some solutions the post-filtering is done alongside the vector search phase, via the inclusion of attribute data in the vector index. 

As PG-based ANNS solutions have performed well in terms of recall and throughput in the \textit{unfiltered} version of the problem, these have been extended for \textit{filtered} ANNS solutions \cite{NEURIPS2023-Wang-NHQ, wang2022navigableproximitygraphdrivennative-NHQ-2, Patel2024ACORN, zhao2022constrainedapproximatesimilaritysearchAIRSHIP, Gollapudi2023FilteredDiskANN}. 
For example, ACORN \cite{Patel2024ACORN} presents a `predicate-agnostic' indexing approach. 
However, the `decomposition-assembly model' \cite{NEURIPS2023-Wang-NHQ} whereby hybrid queries are split into two problems, addressed by different indexing solutions, is difficult to apply to PGs. 
Uncorrelated attribute/vector data may lead to incorrect graph traversal paths; assumptions about query predicates and selectivity may be required; multiple sub-graphs may need to be built to cater for different attributes or assumed filter predicates. 
As a result, filtered PG-based approaches often restrict predicates to only include a single attribute, or only support low-cardinality `tag'-based attributes, with only equality operators catered for. In contrast, SQUASH caters for unrestricted numbers/types of attributes and predicates. 

Bitwise distance comparisons based on the low-bit OSQ index in SQUASH are used to avoid expensive Euclidean query-to-vector distance calculations;
the use of Hamming distances on binary-quantized data enables rapid pruning without compromising accuracy, particularly in constrained environments \cite{MARUKATAT20131101, Martin2015}. Recent work has shown that randomized bit string-based quantization schemes can be effective \cite{RabitQ2024}; further work could apply these techniques in the context of filtered, distributed and serverless search. Another complementary direction is to adapt optimizations in data warehouses and data lakes, such as mechanisms for fine-grained data skipping based on query patterns \cite{2014SunDataSkipping}.
 
While cloud providers offer various server configurations (e.g., CPU, GPU, HPC), these solutions lack dynamic scaling to meet fluctuating demand. Serverless FaaS has been applied to data-intensive tasks such as TPC-H queries \cite{Muller2020Lambada, Perron2020} and ML inference \cite{Oakley2024FSDInference, Gillis2021, Jarachanthan2021AMPS, Oakley2024ForesightPlus}. Several commercial serverless ANNS solutions have been developed \cite{Weaviate, DatastaxAstraDB,Pinecone-Serverless, TurboPuffer, Upstash}.
The only FaaS-based system for vector similarity search, Vexless \cite{Su2024Vexless}, lacks attribute filtering support. 
It utilizes stateful cloud functions to alleviate communication requests, employs HNSW as the indexing solution (making it challenging to extend Vexless to support rich hybrid search functionality), and introduces a workload generator to model large-scale vector search workloads; Vexless then uses these workloads to perform result caching, but its performance is unclear without this advantage.
In contrast, SQUASH leverages synchronous FaaS invocation to achieve communication at high parallelism levels, and offers rich hybrid search support.
% , and does not rely on caching to achieve high levels of throughput.

\section{Experimental Analysis}
\label{s:5-squash-expts}

\subsection{Experimental Setup and Datasets}
We compare SQUASH against two state-of-the-art serverless vector search offerings (a commercial offering and Vexless \cite{Su2024Vexless}), plus two server-based baselines. We explore the performance characteristics at varying parallelism levels across multiple datasets. Finally, we evaluate the performance using our optional caching module targeted at sustained workloads. %, where caching is used to improve throughput.
We run experiments on high-dimensional vector data benchmarks: \textbf{SIFT1M}, \textbf{GIST1M}, \textbf{SIFT10M} and \textbf{DEEP10M} (see Table \ref{tab:datasets}). 
For Local Intrinsic Dimensionality (LID) \cite{fu2021high-LID-Higher}, a higher figure indicates a more challenging dataset. 
% While a large range of similar reference datasets now exist, 
We selected these datasets in line with Vexless \cite{Su2024Vexless}, and added an additional dataset to allow us to further assess the scalability of our approach. 
We use $recall@k = \frac{G \cap R}{k}$, where $G$ is the set of ground truth nearest neighbors (satisfying the filter predicate) and $R$ is the retrieved set.
For all datasets, we use segment size $S=8$, and select a bit budget $b$ = $4 \times d$. 
Our queries have an approximate (joint) attribute selectivity (i.e., proportion of vectors passing all filters) of $8\%$ in line with other works on hybrid search \cite{Patel2024ACORN}.
We achieve this by generating $A=4$ uniform attributes for each dataset.
To illustrate the scalability of SQUASH to large workloads, we consider the processing of \textbf{1000 queries} for all experiments. 
We report the average of 3 runs, and the same queries are used for all solutions. 

% \begin{table*}[h]
\begin{table}[h]
\centering
\caption{Datasets}
\begin{tabular}{|l|c|c|c|c|}
    \hline
     & \textbf{SIFT1M} & \textbf{GIST1M} & \textbf{SIFT10M} & \textbf{DEEP10M} \\ \hline
     $N$ & 1,000,000 & 1,000,000 & 10,000,000 & 10,000,000 \\ \hline
     $d$ & 128 & 960 & 128 & 96 \\ \hline
     % Data Type & int32 & int32 & int32 & float32 \\ \hline
     % LID\cite{li2019approximateLID} & 9.3 & 18.9 & 9.3 & 12.1 \\ \hline
     LID\cite{fu2021high-LID-Higher} & 12.9 & 29.1 & 12.9 & 10.2 \\ \hline
     Bit budget $b$ & 512 & 3840 & 512 & 384  \\ \hline
\end{tabular}
\label{tab:datasets}
% \end{table*}
\end{table}

\subsection{Baselines}
\label{ss:7-baselines}

We first compare against a commercial serverless vector database, which we refer to as \textbf{System-X}.
System-X offers a fully managed, cloud-based solution for high-dimensional vector search, with pay-as-you-use pricing, although it does not run on FaaS. 
To evaluate System-X, we first 
locally transform the data into the required format (i.e., a dictionary for each record with keys: ID, values (vector data), metadata (attribute data)). 
We then `upsert' the data, constituting both the upload and indexing phases. 
We use Python's \textsc{ThreadPoolExecutor} on a large server (Intel Xeon W-2245 CPU, 503GB memory) to send parallel requests to System-X.
SQUASH is also evaluated against \textbf{Vexless}\cite{Su2024Vexless}, the only other serverless solution built using FaaS.
Vexless\footnote{https://github.com/Vexless/Vexless} leverages caching in conjunction with their workload generator, to model large query volumes over long test time periods (e.g., several thousand queries issued per second for an entire day). 
While this is a sensible approach if repeated queries are frequently submitted, the performance figures it enables may not be representative of individual test runs against previously unseen queries. 
Additionally, Vexless does not offer support for hybrid queries, an important feature of SQUASH which requires significant computation in the QueryAllocators (QAs). 
Therefore, we compare with Vexless separately 
in Section \ref{ss:7-caching}. 
In all other experiments, \textit{SQUASH does not use result caching}.
Finally, the cost and performance of SQUASH are compared against several \textbf{server-based baselines}. 
For all server experiments, the same codebase as SQUASH is used, modified to run on a single machine (i.e., spawning separate processes rather than invoking parallel Lambda functions).

\subsection{FaaS and Server Setup}

We use AWS Lambda as the FaaS compute platform. Single Lambda applications are created for the \textit{QA} and \textit{CO} components, described in Section \ref{s:3-serverless}. 
For the \textit{QPs}, a Lambda application is created per partition, e.g., \textit{squash-processor-0, squash-processor-1}.  
We allocate $M_{CO}$ = 512MB to the CO and $M_{QA} = M_{QP} $ = 1770MB to the QA and all QPs; 1770MB has been reported as a cut-off point where an additional core is allocated to the function \cite{AWS-Lambda-1770MB, FN-AWS-Lambda-Memory-Compute}. 
% Boto3 (AWS Python SDK) is used to interact with all AWS services. 
We invoke concurrent query-parallel QAs $N_{QA}$ = 10 ($F=10, l_{max}=1$), 20 ($F=4, l_{max}=2$), 84 ($F=4, l_{max}=3$), 155 ($F=5, l_{max}=3$), 258 ($F=6, l_{max}=3$), 340 ($F=4, l_{max}=4$), each of which can invoke multiple processors. 
Python's \textsc{ThreadPoolExecutor} is employed to parallelize the invocation of child QA/QP functions in Lambda instances. 
SQUASH was built using Python 3.11, NumPy 2.0.0 and Bitarray 2.5.0. $P$ = 10 partitions are used for \textbf{SIFT1M} and \textbf{GIST1M} and $P$ = 20 for \textbf{SIFT10M} and \textbf{DEEP10M}, in accordance with the sizes of the datasets. 
For all experiments, our entire pipeline described in Section \ref{ss:2.4-search-pipeline} is run. 
All experiments are performed in the AWS eu-west-1 region. In all cases, SQUASH is calibrated to achieve the same 97\% recall as System-X, although our solution can achieve $>99\%$ recall if configured to do so.
The parameters used to achieve this are as follows: Binary Quantization Cut-off Percentage $H_{perc}=10$, Fine-Tuning Ratio $R=2$ for all datasets. 
For the Centroid Distance Threshold $T$, we use $\beta = 0.001$. SIFT1M: $T=1.15$, GIST1M: $T=1.2$, SIFT10M: $T=1.15$, DEEP10M: $T=1.13$.
For our server baselines, two instance sizes are selected, enabling us to compare performance/cost against relatively small and large instances. 
AWS EC2 is used for all server baselines. 
Our larger instance type is c7i.16xlarge (64 vCPU, 128GB memory). 
We also baseline against a c7i.4xlarge instance (16 vCPU, 32GB memory) to evaluate a more cost-effective option. 

\begin{figure}[h]%[h!]%
\centering  %
\includegraphics[scale=0.3]{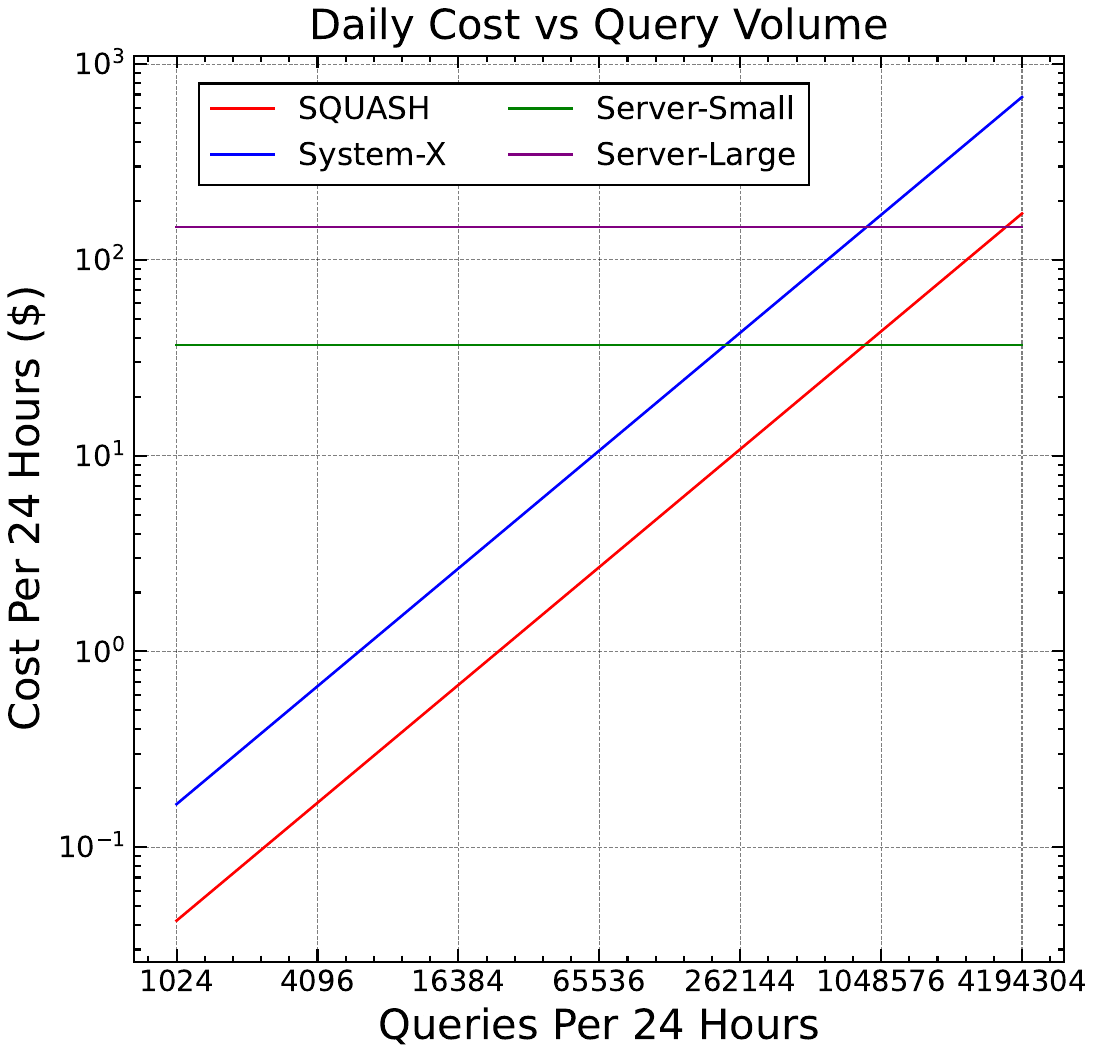}
\caption{Daily cost of SQUASH, System-X and small/large servers for various (uniform) query volumes}
\label{fig:squash-daily-cost-plot}
\end{figure}

\subsection{Cost and Performance Comparison}

\begin{figure}[h]%[h!]%
\centering  %
\includegraphics[scale=0.32]{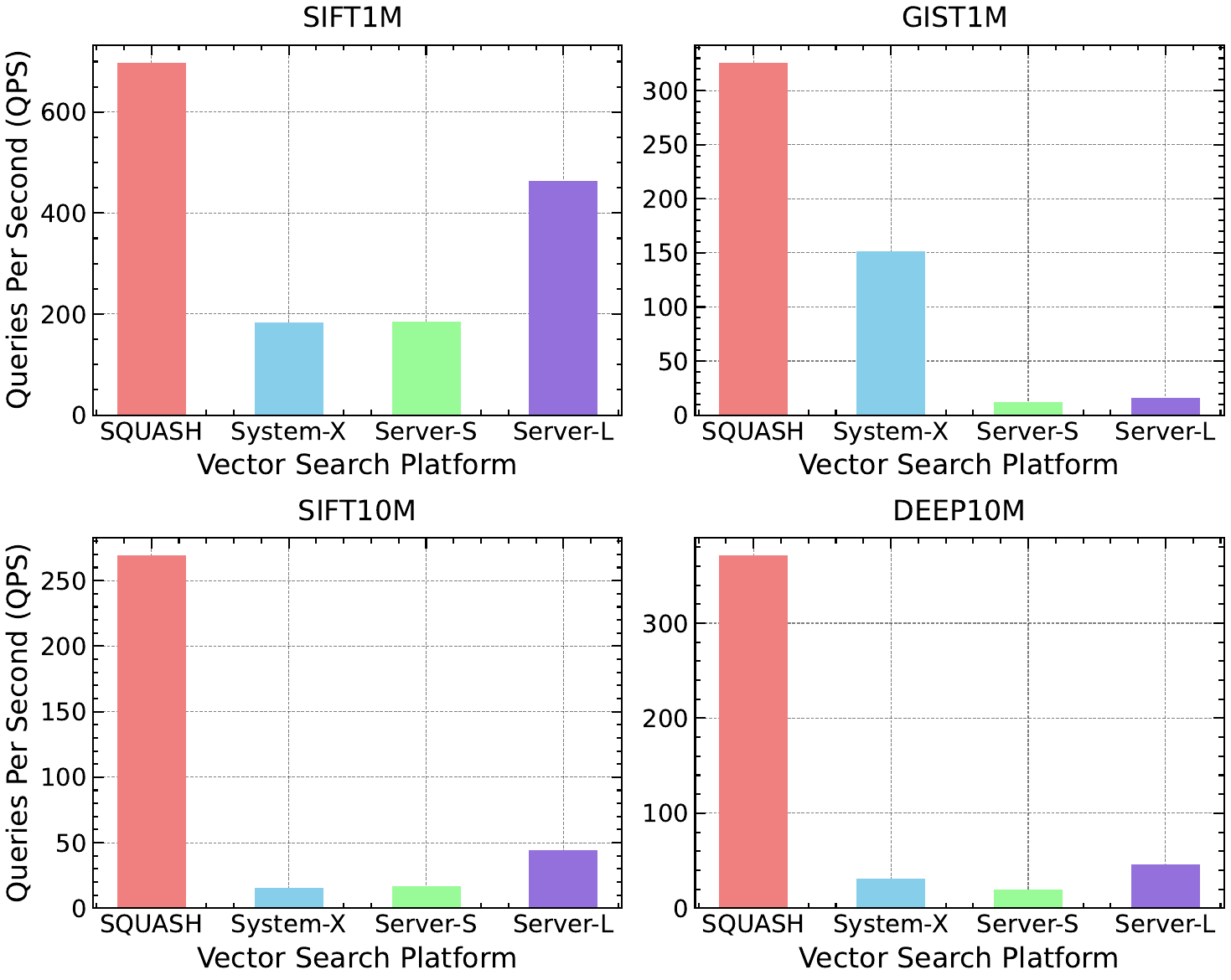}
\caption{Queries per second (QPS) for SQUASH, System-X and server-based baselines}
\label{fig:squash-performance-graph}
\end{figure}

First, the cost-effectiveness of SQUASH is evaluated against System-X and both server-based baselines in Figure \ref{fig:squash-daily-cost-plot}.
In this study, a sporadic vector search workload is modeled, where queries arrive at uniform intervals over a 24 hour period. 
These requests are evenly distributed over the SIFT1M, GIST1M, SIFT10M and DEEP10M datasets. 
For this experiment, a balanced SQUASH configuration is selected ($N_{QA} = 84$), which achieves an attractive cost-to-performance ratio (N.B., this is not the cheapest available, so costs can be reduced further if latency requirements are less strict). 
For System-X, costs are calculated based on the number of read pricing units consumed by each request. 
For the server baselines, it is assumed that two of the respective instances are provisioned, to accommodate periods of bursty traffic and to offer redundancy.
It is observed that \textbf{SQUASH is consistently cheaper per-request than System-X} (per-query cost reductions for each dataset - SIFT1M: 5x, GIST1M: 3.8x, SIFT10M: 3.6x, DEEP10M: 4.1x).
Significant cost savings are also achieved compared to all server-based solutions until very large daily query volumes are reached (\textbf{approx 1M / 3.5M queries per day}).
Next, we compare the queries per second (QPS) processed by each vector search solution in Figure \ref{fig:squash-performance-graph}. 
It is noted that SQUASH achieves significantly higher QPS than System-X for all datasets, with up to an $\sim$18x improvement for SIFT10M. System-X is most competitive on GIST1M, but the significant query parallelism of SQUASH leads to higher throughput. 
Both server baselines struggled with scalability. The nature of the SQUASH system requires high parallelism to function effectively; even on a large server, there is contention between QA and QP processes. In contrast, the ability of FaaS to offer thousands of independent parallel instances gives it a significant scalability and performance advantage.

\subsection{SQUASH Cost-to-Performance Trade-Off}

\begin{figure}[]%[h!]%
\centering  %
\includegraphics[scale=0.4]{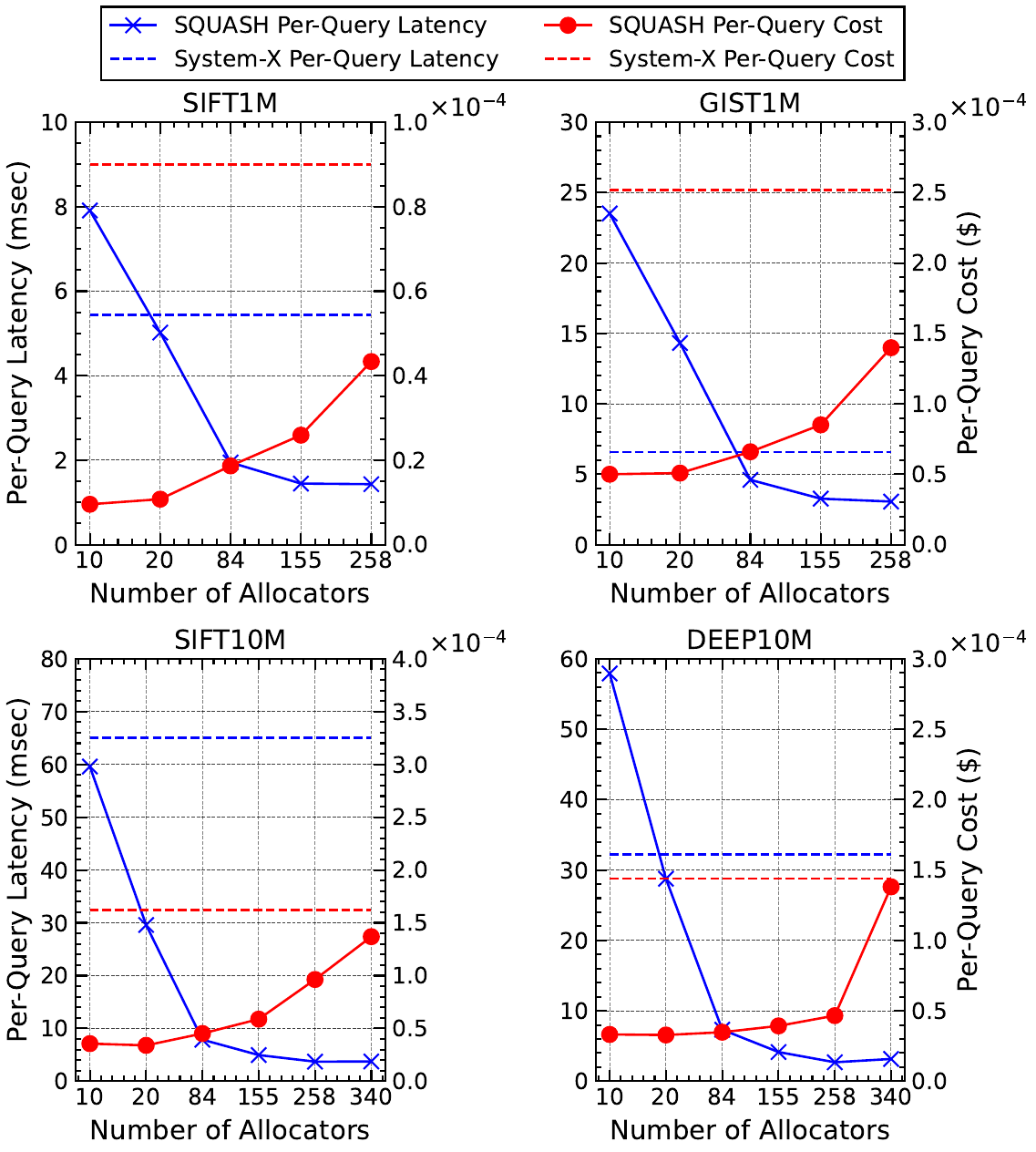}
\caption{Runtime and cost of SQUASH with varying $N_{QA}$}
%, $P = 8, 20, 42$ and $62$}.
\label{fig:squash-latency-and-cost-2x2}
\end{figure}

Next, in Figure \ref{fig:squash-latency-and-cost-2x2} we explore the scalability and cost-to-performance trade-off of our fully serverless solution, by varying the number of QueryAllocators ($N_{QA}$) invoked for each batch query execution. 
Recall that QAs are responsible for performing attribute filtering, as well as cluster ranking/selection; therefore, the total number of QPs invoked scales with $N_{QA}$, up to thousands of concurrent QPs for large $N_{QA}$. 
In Figure \ref{fig:squash-latency-and-cost-2x2}, the dotted horizontal lines correspond to the System-X latency/cost for each dataset; SQUASH achieves lower latency at all but the lowest parallelism levels, and lower costs at all parallelism levels. 
For GIST1M, where System-X offers the most competitive latency compared to SQUASH with $N_{QA}=84$, we see significant cost savings with SQUASH.
It is observed that invoking $N_{QA} = \{84, 155\}$ offers an attractive balance of cost and performance, particularly for SIFT1M and GIST1M. 
At these parallelism levels, each QA is responsible for managing $\sim$ 12/7 queries respectively in each batch. 
Using additional QAs for these datasets may yield slightly higher QPS, albeit it with a disproportionate cost increase.
It should be noted that GIST1M latencies are $\sim 2.3$x higher than those of SIFT1M, reflecting the challenge of increasing the dimensionality by a factor of almost 8.
For the larger SIFT10M and DEEP10M datasets, our results indicate that while $N_{QA}=84$ still offers a reasonable balance of cost and performance, additional throughput can be achieved at higher parallelism levels, up to $N_{QA}=258$. 
It is noted that $N_{QA}=340$ is generally not recommended for this workload, as the overhead of invoking the thousands of concurrent FaaS instances this produces (even accelerated by our tree-based invocation method and warm starts) dominates the time spent performing computation. A larger query set would be required for this to be beneficial.

% \begin{table*}[h]
\begin{table}[h]
\centering
\caption{Performance with caching, recall=0.97}
\begin{tabular}{|l|c|c|c|}
    \hline
     & \textbf{GIST1M} & \textbf{SIFT10M} & \textbf{DEEP10M} \\ \hline
     Vexless\cite{Su2024Vexless} QPS & 285 & 3125 & 2500 \\ \hline
     SQUASH QPS & \textbf{326} & \textbf{3388} & \textbf{2804} \\ \hline
     SQUASH Cache Ratio & \textbf{1} & 10 & 8 \\ \hline
\end{tabular}
\label{tab:caching-performance}
% \end{table*}
\end{table}
\subsection{Performance with Caching}
\label{ss:7-caching}

As discussed in Section \ref{ss:7-baselines}, Vexless \cite{Su2024Vexless} employs a caching solution to improve query performance, which is a valid approach for sustained workloads. 
It is also evaluated over long periods of high query traffic (e.g., 3k+ queries per second for an entire day, over 250M+ in total), repeating the same 1k/10k reference queries. % provided with the benchmark datasets. 
However, this induces significant query repetition and thus many cache hits, so the reported latencies may not be reflective of execution against previously unseen queries.
To allow us to evaluate SQUASH in a similar context, here we also leverage a lightweight caching solution. 
For this study, we seek to determine the `cache ratio' (i.e., the number of times we duplicate the same queries) required for SQUASH to achieve higher QPS than Vexless, on each common dataset. We use the same recall target as in our other experiments for both systems. These figures are shown in Table \ref{tab:caching-performance}. For GIST1M ($d = 960$), SQUASH is able to achieve higher throughput with no query duplication; this illustrates the impressive performance of our SQ-based approach.% 
\section{Conclusion}
\label{s:6-squash-conclusion}

In this paper, we present SQUASH, the first fully serverless high-dimensional hybrid vector search solution. We introduce OSQ, a highly parallelizable quantization-based approach for both vectors and attributes. 
OSQ combines the vector approximations of multiple dimensions into shared segments, significantly increasing compression.
SQUASH leverages a multi-stage search pipeline, including attribute filtering, partition ranking and selection, effective pruning via low-bit OSQ, a minimal set of distance calculations and an optional post-refinement step. 
This workflow is designed to minimize the load on compact FaaS instances. 
SQUASH introduces several novel elements for fully serverless vector search, including a hierarchical tree-based invocation scheme which enables the rapid launch of thousands of concurrent instances. 
SQUASH also introduces data retention exploitation (DRE) in serverless systems, which eliminates redundant I/O, reduces costs and significantly improves performance. Detailed experiments on several large vector data benchmarks confirm that SQUASH achieves significant scalability and cost-effectiveness, with up to 18x higher throughput and 9x cost savings compared to state-of-the-art serverless vector search offerings. 

\begin{acks}
This work was supported in part by the Feuer International Scholarship in Artificial Intelligence.
\end{acks}

%\clearpage

\bibliographystyle{ACM-Reference-Format}
\bibliography{sample}

\end{document}